\documentstyle[graphics]{mn}

\newcommand{\etal}{{et al.}~}

\newcommand{\bc}{\begin{center}}
\newcommand{\be}{\begin{equation}}
\newcommand{\ee}{\end{equation}}
\newcommand{\ec}{\end{center}}

\newcommand{\hydra}{{\sc{HYDRA }}}

\newcommand{\cmbfast}{{\sc {CMBFAST }}}
\newcommand{\vpfit}{{\sc {VPFIT }}}
\newcommand{\autovp}{{\sc {AUTOVP }}}
\newcommand{\fitlyman}{{\sc {FITLYMAN }}}
\newcommand{\lya}{\mbox{Ly$\alpha$~}}

\renewcommand{\H}{{\mbox{${\rm H{\sc i}~}$}}}

\newcommand{\He}{{\mbox{${\rm He{\sc i}~}$}}}
\newcommand{\Hep}{{\mbox{${\rm He{\sc ii}~}$}}}
\newcommand{\Hepp}{{\mbox{${\rm He{\sc iii}~}$}}}

\newcommand{\ltsima}{\mbox{$\; \buildrel < \over \sim \;$}}
\def \simlt{\lower.5ex\hbox{\ltsima}}            
\def \gtsima{\mbox{$\; \buildrel > \over \sim \;$}}
\def \simgt{\lower.5ex\hbox{\gtsima}}            

\newcommand{\mnras}{MNRAS}
\newcommand{\apj}{ApJ}
\newcommand{\apjl}{ApJL}
\newcommand{\pasp}{PASP}
\newcommand{\nat}{Nature}
\newcommand{\aap}{A\&A}
\newcommand{\aj}{AJ}
\title[Temperature fluctuations in the IGM] {Temperature fluctuations
in the intergalactic medium\thanks{The data presented herein were
obtained at the W.M.Keck Observatory, which is operated as a scientific
partnership among the California Institute of Technology, the
University of California and the National Aeronautics and Space
Administration. The Observatory was made possible by the generous
financial support of the W.M.Keck foundation.}\thanks{Based on public
data released from the VLT/UVES Commissioning and Science
Verification and from the OPC porgramme 65.O-296A (PI S.-D'Odorico) at
the VLT/Kueyen telescope, ESO, Paranal, Chile.}}

\author[Theuns et al.]{Tom Theuns$^{1}$, Saleem
Zaroubi$^{2}$, Tae-Sun Kim$^{3}$, Panayiotis Tzanavaris$^{1}$,\and
and Robert F. Carswell$^{1}$\\ $^{\,1}$ Institute of Astronomy, Madingley Road,
Cambridge CB3 0HA, UK\\ $^{\,2}$ Max-Planck Institut f\"ur Astrophysik,
Postfach 123, 85740 Garching, Germany\\ $^{\,3}$ European Southern
Observatory, Karl-Schwarzschild-Stra\ss e 2, D-85748 Garching bei
M\"unchen, Germany}

\begin{document}
\maketitle

\begin{abstract}
The temperature of the low-density intergalactic medium is set by the
balance between adiabatic cooling resulting from the expansion of the
universe, and photo-heating by the UV-background. We have analysed the
\lya forest of eleven high-resolution quasar spectra using wavelets,
and find strong evidence of a marked jump in the temperature at the
mean density, $T_0$, of 60 $\pm$ 14 per cent over the redshift interval
$z=[3.5, 3.1]$, which we attribute to reionization of \Hep. The jump
can be seen in all three of our spectra that straddle redshift 3.3, at
a significance of $\ge 99$ per cent. Below $z\sim 3.1$, our results are
consistent with a smooth cooling down of the universe, as expected when
adiabatic expansion dominates over photo-heating by a UV-background
from QSOs and galaxies. We find no evidence of thermal fluctuations on
scales $\ge 5000$ km s$^{-1}$ larger than 50 per cent, which could be
detected by our method, suggesting that the IGM follows a reasonably
well-defined temperature-density relation. We demonstrate that the mean
wavelet amplitude $\langle A\rangle\propto 1/T_0$ and calibrate the
relation with hydrodynamical simulations. We find $T_0\ge 1.2\times
10^4$K at $z\ge 3.6$.  Such high temperature suggest that \H
reionization occured relatively recent.
\end{abstract}


\section{Introduction}
The intergalactic medium (IGM) can be observed in the spectra of
distant objects such as quasars through resonant absorption in the
Lyman-$\alpha$ transition of neutral hydrogen (Bahcall \& Salpeter
1965; Gunn \& Peterson 1965). The availability of high resolution
echelle spectrographs on large telescopes (HIRES on Keck \& UVES on
VLT) has provided us with data of unprecedented quality over recent
years (see Rauch 1998 for a recent review). At the same time, a
theoretical paradigm within the context of the cold dark matter
cosmology has emerged (e.g. Bi, Boerner \& Chu 1992), well tested by
numerical hydrodynamical simulations(Cen \etal 1994; Zhang, Anninos \&
Norman 1995; Miralda-Escud\'e \etal 1996; Hernquist \etal 1996; Wadsley
\& Bond 1996; Zhang \etal 1997; Theuns \etal 1998; Machacek et al 2000)
, in which the absorption is produced by volume filling, photo-ionised
gas, that contains most of the baryons at redshifts $z\sim 3$ (see
e.g. Efstathiou, Schaye \& Theuns 2000 for a recent review). The
absorbers are locally over dense extended structures, close to local
hydrostatic equilibrium (Schaye 2001). The combination of a predictive
theory and superb data has led to a veritable revolution in IGM
studies.

On large scales, the gas distribution is similar to that of the
underlying dark matter, but on small scales pressure forces smooth the
gas distribution, erasing most of the small scale power. Combined with
thermal broadening and peculiar velocities, this results in a strong
suppression in the amplitude of the flux power spectrum on scales $\le
200$ km s$^{-1}$ (e.g. Theuns, Schaye \& Haehnelt 2000). Another
consequence of the thermal content of the gas is a cut-off in the
distribution of line widths $b$ below $\sim 20$ km s$^{-1}$ (Schaye et
al 1999) , which is clearly present in observed line samples as well
(e.g. Kirkman \& Tytler 1997). Both these signatures can be used to
infer the IGM temperature by calibrating the cut-off using simulations
(Schaye et al 1999, 2000; Ricotti, Gnedin \& Shull 2000; Bryan \&
Machacek 2000; McDonald et al 2000).

The IGM temperature can be used to constrain the reionization history
because the thermal time scales are long in the low density gas probed
by the \lya absorption (Miralda-Escud\'e \& Rees 1994). Consequently
the gas retains a memory of its heating {\em history}. At over
densities $\delta\rho/\langle\rho\rangle\le 3$, shock heating and
radiative cooling are unimportant and photo-heating of the expanding
gas introduces a tight density-temperature relation
$T/T_0=(\rho/\langle\rho\rangle)^{\gamma-1}$ (e.g. Hui and Gnedin
1997). A sudden epoch of reionization will make the gas nearly
isothermal ($\gamma\sim 1$) at a temperature of $\sim 10^4$K.  Away
from reionization, the density-temperature relation steepens again,
asymptotically reaching $\gamma\sim 1.6$. The steepening occurs because
denser gas has a higher neutral fraction, and hence a larger
photo-heating rate. If the heating rate can be computed reliably, given
the observed sources of ionising photons, then a measurement of the
entropy of the gas can in principle determine the reionization
epoch. If the sources responsible for reionizing hydrogen have a soft
spectrum, there may be more than one reionization epoch since helium
reionization requires harder photons.

There is some observational evidence that \Hep\\ reionization occurs
around $z\sim 3$, based on the occurrence of large fluctuations in the
\Hep optical depth (Davidsen et al. 1996; Reimers et al. 1997; Heap et
al. 2000), and the change in hardness of the ionising spectrum as
inferred from metal-line ratios (Songaila \& Cowie 1996; Songaila
1998). The interpretation of the optical depth data is not
straightforward however (Miralda-\'Escude et al. 2000), and there is
some discussion in the literature about the reality and interpretation
of the claimed change in metal-line ratios (Boksenberg, Sargent \&
Rauch 1998; Giroux \& Shull 1997). Note that, if \Hep reionization
does not occur instantaneously, various measures of the \Hep abundance
change may well find a range of \lq reionization epochs\rq, given that
they may sample different over densities.

Reionization is not expected to be instantaneous because ionisation
fronts expand much faster into low density voids than into the higher
density filaments (e.g. Gnedin 2000a). If the sources of ionising
photons are bright but scarce, then reionization will not be complete
until the large ionised bubbles overlap. A consequence of such patchy
reionization is that different regions might follow different $\rho-T$
relations and so have different values of $T_0$. Such \lq temperature
fluctuations\rq\ may lead to apparent discrepancies between the
different algorithms discussed earlier to measure $T_0$ and $\gamma$,
since different methods may give other weights to regions in the same
spectrum.

The aim of this paper is to test whether the data are consistent with a
single well defined $\rho-T$ relation and to investigate whether \Hep
reionization can be detected by a jump in temperature. We use the
wavelet analysis proposed by Theuns \& Zaroubi (2000), and apply it to
a set of high resolution QSO spectra spanning a wide range of
redshifts. We use hydrodynamical simulations to demonstrate that our
method can detect variations in the amplitude $T_0$ of the
temperature-density relation $\rho/\langle\rho\rangle\propto
(T/T_0)^{\gamma-1}$ of order 50 per cent, even when these occur on
scales as small as $\sim 5000$ km s$^{-1}$ (1000km s$^{-1}$ corresponds
to $\sim 9.3$ co-moving Mpc h$^{-1}$ and a redshift extent $\delta
z=1.33\times 10^{-2}$ at redshift $z=3$, in the currently popular flat,
cosmological constant dominated model with matter density
$\Omega_m=0.3$.) The method presented here does not use simulations to
identify regions of different temperatures.  Recently, Zaldarriaga
(2001) applied a similar analysis to QSO 1422+231 and showed that a
model with two temperatures, that occupy comparable fractions of the
spectrum, is constrained to have temperature variations smaller than a
factor of 2.5.

This paper is organised as follows. Section~2 presents the data and
Section~3 the simulations used to demonstrate the method. Section~4
describes the wavelet decomposition and the statistical tools making
use of hydrodynamical simulations to illustrate the approach. Section~5
presents the results, and these are discussed in Section~6. Section~7
summarises. Readers not interested in the details of the method can
skip directly to section~4.3.1, which provides an illustration of the
method for a simulated spectrum with imposed temperature fluctuations.

\section{The data}

In our analysis, we have used data obtained with the high resolution
spectrograph on the Keck I telescope (Vogt et al. 1994), and UVES, the
ultraviolet echelle spectrograph on the VLT (Kueyen) telescope (D'
Odorico et al. 2000). We have combined line lists taken from the
literature with publically available spectra (Table\ref{table:data}),
to obtain a set of eleven high resolution QSO spectra that span a wide
range of emission redshifts $z_{\rm em}= 1.7\rightarrow 3.7$. High
resolution data is required since the lines are intrinsically narrow
$\sim 20$ km s$^{-1}$. The standard data reduction is described in the
original references, with the exception of APM~0827+5255 which was
reanalyzed after calibration of the flux scale to remove echelle order
mismatches (Tzanavaris \& Carswell 2002). All spectra have signal to
noise of 40--50 per resolution element, and a similar resolution, $R
\sim 40,000$. Only two out of eleven of these spectra (QSOs
APM~0827+5255 and 1422+293) were used in the analysis of the thermal
evolution by Schaye et al (2000).

Line profile fitting has been performed on the data. Voigt profiles are
fitted to the absorption lines using a $\chi^2$ minimisation procedure,
producing a list of lines with given column density, $N_\H$
(cm$^{-2}$), width $b$ (km s$^{-1}$) and absorption redshift $z$. Where
possible, metal lines have been identified from line coincidences. This
is in fact a crucial step, since such lines tend to be narrow and so
might be mistaken for a cold \lya cloud.

For QSOs Q0636+680, Q0956+122 and Q0014+813, we use the published line
list given by Hu et al (1995), who used their own automated line
fitting programme. All other spectra have been analysed with the
semi-automatic line-fitting programme VPFIT\footnote{for details on
VPFIT, see\\ http://www.ast.cam.ac.uk/\~{}rfc/vpfit.html} (Webb 1987;
Carswell et al. 1987). The mock spectra from our simulations, described
in the next section, are also fitted using VPFIT.

The line parameters from Voigt profile fitting are not unique (Kirkman
\& Tytler 1997). However, in the analysis described below, we never use
the detailed properties of the lines, but rather analyse spectra as
reconstructed from the line list. For QSO HE~1122--1648, we have redone
our analysis using another line fitting programme, \fitlyman (Fontana \&
Ballester 1995), and the results are nearly identical.

\begin{table*}
\caption[]{QSO spectra used}
\label{tab1}
\begin{tabular}{lccccll}
\hline
\noalign{\smallskip}
QSO & $z_{\mathrm{em}}$ & $B^{\mathrm{a}}$ & $\lambda\lambda$ &
$z_{\mathrm{Ly\alpha}}$ & Comments & Ref. \\
\noalign{\smallskip}
\hline
HE0515--4414  & 1.719 & 14.9 & 3080--3270 & 1.53--1.69 & VLT/UVES   & 1 \\
J2233--606    & 2.238 & 17.5 & 3400--3890 & 1.80--2.20 & VLT/UVES   & 1,2 \\
HE1122--1648  & 2.400 & 17.7 & 3500--4091 & 1.88--2.37 & VLT/UVES   & 2 \\
HE2217--2818  & 2.413 & 16.0 & 3510--4100 & 1.89--2.37 & VLT/UVES   & 1,2 \\
Q0636+680     & 3.174 & 16.5 & 4300--4900 & 2.54--3.03 & Keck/HIRES & 3 \\  
Q0302--003    & 3.286 & 18.4 & 4808--5150 & 2.96--3.24 & VLT/UVES   & 2 \\
Q0956+122     & 3.301 & 17.8 & 4400--5000 & 2.62--3.11 & Keck/HIRES & 3 \\
Q0014+813     & 3.384 & 16.5 & 4500--5100 & 2.70--3.20 & Keck/HIRES & 3 \\
Q1422+231     & 3.620 & 16.5 & 3645--7306 & 2.91-- 3.60& Keck/HIRES & 4 \\
Q0055--269    & 3.655 & 17.9 & 4852--5598 & 2.99--3.60 & VLT/UVES   & 2 \\
APM08279+5255 & 3.911 & 15.2 & 4400--9250 & 3.20--3.72 & Keck/HIRES  & 5 \\ 
\noalign{\smallskip}
\hline
\label{table:data}
\end{tabular}
\begin{list}{}{}
\item[$^{\mathrm{a}}$] $B$-band magnitudes from the SIMBAD astronomical database,
except for QSO APM08279+5255, for which the $R$-band magnitude is from  Irwin et al. (1998). 
\item[] References: 1. Kim et al. (2001a); 2. Kim et al. (2001b); 3.
Hu et al. (1995); 4. Rauch et al. (1997); 5. Ellison et al. (1999)
\end{list}
\end{table*}

\section{Hydrodynamical simulations and mock spectra}
\label{sect:simulations}
We use hydrodynamical simulations to illustrate the method described
below. Briefly, these are simulations of a flat, vacuum-energy
dominated cold dark matter model (matter density $\Omega_m=0.3$, baryon
fraction $\Omega_b h^2=0.019$, Hubble constant $H_0=100h$ km s$^{-1}$
Mpc$^{-1}$, $h=0.65$ and normalisation $\sigma_8=0.9$. We have used
\cmbfast (Seljak \& Zaldarriaga 1996) to compute the appropriate linear
transfer function.

The simulation code is based on \hydra (Couchman, Thomas \& Pearce
1995) and combines Smoothed Particle Hydrodynamics (SPH, Gingold \&
Monaghan 1977; Lucy 1977) with adaptive P3M gravity (Couchman 1991). It
has been modified extensively by one of us (TT) in order to be able to
simulate the IGM. The code has been tested comprehensively, and has
been parallellized in the OpenMP standard. Non-equilibrium radiative
processes (cooling, photo-ionization heating by a UV-background and
Compton cooling off the cosmic micro-wave background radiation) are
included with rates given as in Theuns et al. (1998). The evolution of
the UV-background is chosen to give the thermal evolution as determined
by Schaye et al. (2000, the simulation is referred to there as the \lq
designer simulation\rq). In this simulation, \H and \He reionize at a
redshift $z\sim 7$ by soft sources, whereas \Hep reionization is
delayed to a redshift $z\sim 3.4$. The simulation box is 12.5$h^{-1}$
co-moving Mpc on a side (corresponding to $\approx 1400$ km s$^{-1}$ at
redshift $z=3$), and gas and dark matter are represented with $2\times
256^3$ particles of masses 1.45 and $8.25\times 10^6M_\odot$
respectively. The resolution of this simulation is sufficient to
resolve the various line broadening mechanisms (Theuns et al. 1998),
which is of course crucial for this type of analysis. At the same time,
effects of missing large scale power on the statistics analysed here,
are not very important at sufficiently high redshifts $\ge 2$ (Theuns
\& Zaroubi 2000).

In order to study the effect of temperature fluctuations on the
properties of the absorption lines produced, we impose specific
temperature-density relations,
$T/T_0=(\rho/\langle\rho\rangle)^{\gamma-1}$, on our simulation
outputs.  We impose such a $T-\rho$ relation on all gas particles with
$\rho\le \langle\rho\rangle$ or $T(\rho) \le 2\times T_0
(\rho/\langle\rho\rangle)^{\gamma-1}$ and $\rho\le
10\langle\rho\rangle$. Models C and H (for Cold and Hot respectively)
have $T_0=1.5\times 10^4$K and $2.2\times 10^4$K respectively, with the
same slope, $\gamma=5/3$. We have used these two models to make mock
spectra of length $\sim 50000$ km s$^{-1}$, typical of observed \lya
spectra of redshift $\sim 3$ QSOs. (These combined spectra use sets of
Voigt profiles pertaining to fits of spectra of individual sightlines
through the simulation box. Each spectrum is fitted such that the
maximum transmission occurs at the edges, by taking advantage of the
fact that the individual chuncks are periodic.) In order to make the
comparison to data more realistic, we impose the observed mean
absorption on the spectra. In addition, we also add noise and \lq
instrumental broadening\rq\, following the procedure described in
detail in Theuns, Schaye \& Haehnelt (2000).

Below we analyse several such mock spectra
(Table~\ref{table:mock}). Spectrum S1 is composed of spectra drawn from
model C for its first half, and from model H for its second
half. Spectrum S2 is drawn from model H, except for a cold gap of
length 5000 km s$^{-1}$ drawn from model C in the middle of the
spectrum. Finally, S3 is a single temperature model (H), but the mean
absorption is scaled to be respectively 0.64 and 0.6 in the lower and
upper redshift halves of the spectrum respectively.

\begin{table}
\caption[]{Mock spectra used. ($T_0,\gamma$) refers to the imposed
temperature-density relation, $V=5\times 10^5$ km s$^{-1}$ is the
length of each spectrum.}
\begin{tabular}{lc}
\hline
\noalign{\smallskip}
\hline
\noalign{\smallskip}
Spectrum & ($T_0,\gamma$) \\
\hline
S1       & ($1.5\times 10^4,5/3$) for $v\le 0.5 V$ \\
         & ($2.2\times 10^4,5/3$) for $v\ge 0.5 V$ \\
S2		   & ($2.2\times 10^4,5/3$) for $|v-V/2| \ge 2500$\\
         & ($1.5\times 10^4,5/3$) else\\
S3       & ($2.2\times 10^4,5/3$)$^1$\\
\noalign{\smallskip}
\hline
\end{tabular}
\begin{list}{}{}
\item[$^{\mathrm{1}}$] Mean  flux $\bar F=0.64$ (0.6) for $v\le V/2$
($v\ge V/2$).
\end{list}
\label{table:mock}
\end{table}

To illustrate the effect of varying $\gamma$, we also compute spectra
for models with imposed $T-\rho$ relations of $(T_0,\gamma)=(1.5\times
10^4,1)$ and $(2.2\times 10^4,1)$. (The neglect of extra broadening due
to pressure effects and peculair velocities in spectra with imposed
equation of state artificially decreases the differences in line shapes
for different $T_0$ models.) In addition to these simulations, we use
another set of four simulations evolved with $2\times 64^3$ particles
in a box of size 2.5$h^{-1}$ Mpc and the same $\Lambda$CDM cosmology as
the others. These simulations have been run with different heating
rates, to give them different values of log$_{10}
T_0=[4.0,4.1,4.2,4.3]$ but the same value of $\gamma\approx 1.4$. The
mock spectra from these simulations are computed in the same way as the
others.

These mock spectra are then scaled to have the same mean absorption as
the data. We have used VPFIT to produce line lists for all mock
spectra, and this allows us to treat data and simulation in an
identical fashion for the wavelet analysis presented in the next
section.

\section{Method}
\subsection{Wavelet projection}
\label{sect:wvlt}
A discrete wavelet is a localised function with a finite bandwidth (see
e.g. Press et al. 1992 for an introduction and original references to
the application of wavelets to a wide variety of problems). This makes
wavelets useful for characterising line widths in a spectrum, since the
amplitude of the wavelet will be related to the width of the line, and
the position of the wavelet to the position of the line. Pando \& Fang
(1996) used a wavelet analysis to describe clustering of \lya
absorption lines. Theuns \& Zaroubi (2000) used the Daubechies 20
wavelet (Daubechies 1988) to characterise temperature fluctuations in a
\lya spectrum. Independently, Meiksin (2000) used the same wavelet
basis to show that most of the information in a spectrum is carried by
a small fraction of the wavelets. More recently, Jamkhedar, Bi and Fang
(2001) used a multi-scale analysis based on wavelets to better describe
the transmission power spectrum taking into account uncertainties in
the applied continuum.

Discrete wavelets are also a set of orthogonal basis
functions. Consequently, we can borrow notation from quantum mechanics
to write any function $|\psi\rangle$ as a sum over its wavelet
projections, $|\psi\rangle=\sum_n
\langle\psi_n|\psi\rangle\,|\psi_n\rangle$.  Here, $|\psi_n\rangle$ is
the wavelet function of level $n$, $\langle\psi_n|\psi\rangle$ the
projection of $|\psi\rangle$ onto $|\psi_n\rangle$, and $n$ the \lq
quantum number\rq\ of the basis function. Characteristic for wavelets
is that they have a finite extent in both the spatial direction, and
the frequency domain, i.e. they are like a wave-packet instead of a
single momentum wave. So we'll need two indices to denote a given
wavelet basis function, $|\psi_{n,p}\rangle$, where $n$ denotes the
characteristic frequency (\lq width\rq), and $p$ the position of the
wavelet.

In this paper we will investigate the properties of the spectrum in
terms of a given single {\em frequency} of the wavelet, i.e. we will
study
\begin{eqnarray}
|\psi(n)\rangle &\equiv& \sum_p
\langle\psi_{n,p}|\psi\rangle\,|\psi_{n,p}\rangle\,.
\label{eq:wvlt}
\end{eqnarray}
Note that summing $|\psi(n)\rangle$ over $n$ will give back the
original function $|\psi\rangle$. We are left to tune the frequency $n$
to make it most sensitive to the temperature. We have chosen to use the
Daubechies 20 wavelet, as in Theuns \& Zaroubi (2000), who give
examples of the decomposition of a spectrum in terms of that
wavelet. Note that computing the wavelet projection Eq.~(\ref{eq:wvlt})
is computationally very similar to performing a fast Fourier transform,
and many standard computer packages have wavelet transforms built-in.
The usage of wavelets is not very common in astronomy, we would
therefore like to note that the computation of the projection
Eq.~(\ref{eq:wvlt}) requires a handful of programming lines in standard
software packages such as {\sc IDL}, and much less than a second of
computer time to evaluate. For a given choice of the wavelet, the
decomposition Eq.~(\ref{eq:wvlt}) is unique.

Theuns \& Zaroubi (2000) showed that the projection in
Eq.~(\ref{eq:wvlt}) is already very sensitive to the temperature of the
gas, because narrow lines tend to generate larger wavelet coefficients
than broader lines, for a suitable choice of $n$. Here we describe four
improvements to that method.

\noindent (1) Since the $|\psi_{n,p}\rangle$ are orthogonal basis
functions, they change sign at least once, in contrast to an absorption
spectrum. In particular, the Daubechies 20 wavelet does not look like
an absorption line at all, but is more similar to its {\em
derivative}. So we found it advantageous to project the scaled {\em
derivative} of the spectrum $F(\lambda)$ with respect to velocity $v$,
$\partial F/\partial v (F+\eta)^{-1}$. Here, the \lq velocity\rq\ $v$ is
defined as a function of wavelength $\lambda$ through
$d\lambda/\lambda=dv/c$, where $c$ is the speed of light, and the
parameter $\eta=0.2$ is introduced to avoid division by zero close to
the zero level.

\noindent (2) The positional localisation of the narrow wavelet (width
$\sim 15$ km s$^{-1}$) that we will use below is far better than what
we require to determine a jump in temperature. We take advantage of
this to improve the frequency resolution -- and hence the sensitivity
to small changes in the line widths -- of the wavelet projection at the
expense of its localisation by computing
\begin{eqnarray}
|\psi(n,\delta)\rangle &\equiv& \sum_p max_{(-1/2,0,1/2)\times\delta}( |
 \langle\psi(n,p)|\psi(-\delta)\rangle
\nonumber\\
&\times & |\psi(n,p+\delta)\rangle|)\,.
\label{eqn:shift}
\end{eqnarray}
So we shift the input function by $\delta$ to the left,
$|\psi(-\delta)\rangle$, project it onto the wavelet, and shift the
result by the same $\delta$ to the right. We do this for
$\delta=-1/2,0,1/2$ times the wavelet's width, and at each position
take the maximum absolute value of the three projections, and denote
the result by $|\psi(n,\delta)\rangle$. This step is not crucial to the
method, but does improve its sensitivity in finding temperature
fluctuations imposed in the mock spectra.

\noindent (3) Observed spectra contain metal lines in addition to \lya
lines. As described in the data section, we use \vpfit to fit Voigt
profiles, $V(b_i,N_i,\lambda_i)$, to the spectrum. Here,
$V(b_i,N_i,\lambda_i,X_i)$ denotes a Voigt profile of species $X_i$
with column density $N_i$, width $b_i$ and centred on $\lambda_i$. Using
this decomposition, we can reconstruct an (almost) \lq metal free\rq\
spectrum,
\begin{equation}
F_{\rm VP}=exp(-\sum_i V(b_i,N_i,\lambda_i,\lya))\,.
\label{eqn:vpfit}
\end{equation}
(Note that we still impose instrumental broadening on this
reconstruction.)

The decomposition Eq.~(\ref{eqn:shift}) is very good at identifying very
narrow lines $\le 15$ km s$^{-1}$. When applying the method to data, it
is sometimes impossible to decide whether such a narrow line is a metal
line or not, because the corresponding other metal transitions may not
fall in the observed region of the spectrum, or fall on top of a strong
\lya line. In cases where a strong wavelet signal results from a single
Voigt profile, we have decided to flag the line as a potential metal
line, and not consider it any further in the rest of the
analysis. Typically only a hand full of lines (out of several hundreds)
are removed per spectrum.  This is of course a somewhat subjective
procedure, but is unlikely to introduce any {\em positive} detection of
a spurious correlation signal.

Using $F_{\rm VP}$ instead of the original spectrum $F$ has the added
benefit that it is also noise free, making it easier to take the
derivative (step 1), and will also be extremely useful in the
statistical analysis. The decomposition in Voigt profiles is not
unique, but since we only use the reconstructed spectrum, and not the
detailed properties of the fitted lines, we expect the spectrum $F_{\rm
VP}$ to be nearly independent of the actual manner in which Voigt
profiles were fitted. In particular, we were unable to detect any
systematic differences in line shapes between the original spectrum and
the reconstruction, using either our mock spectra or observed data.  We
also redid the analysis for one QSO spectrum (QSO HE~1122--1648) using
\fitlyman (Fontana \& Ballester 1995), another line fitting programme,
and found the same results.

In the absence of evolution in the $b$ parameters, line widths
$\Delta\lambda$ will increase with redshift $\propto (1+z)$. This pure
\lq expansion\rq~ effect can be compensated for by rebinning the
spectrum to velocity space, and \vpfit performs such a rebinning.

\noindent (4) The spectrum also contains strong lines. Their square
shape tends to generate large wavelet amplitudes, an effect similar to
the Gibbs phenomenon familiar from Fourier analysis. In addition, in
the saturated region of the strong lines, the wavelet amplitude becomes
zero. These fluctuations in the wavelet amplitudes do not contain any
information on the IGM temperature and so we ignore them. We begin by
identifying all regions of the spectrum within 1 per cent of being
black. All pixels within 1.5 times the wavelet-width of such a region
are ignored in the statistical analysis described in the next section.

\begin{figure*}
\setlength{\unitlength}{1cm} \centering
\begin{picture}(7,12)
\put(-5., 0){\includegraphics{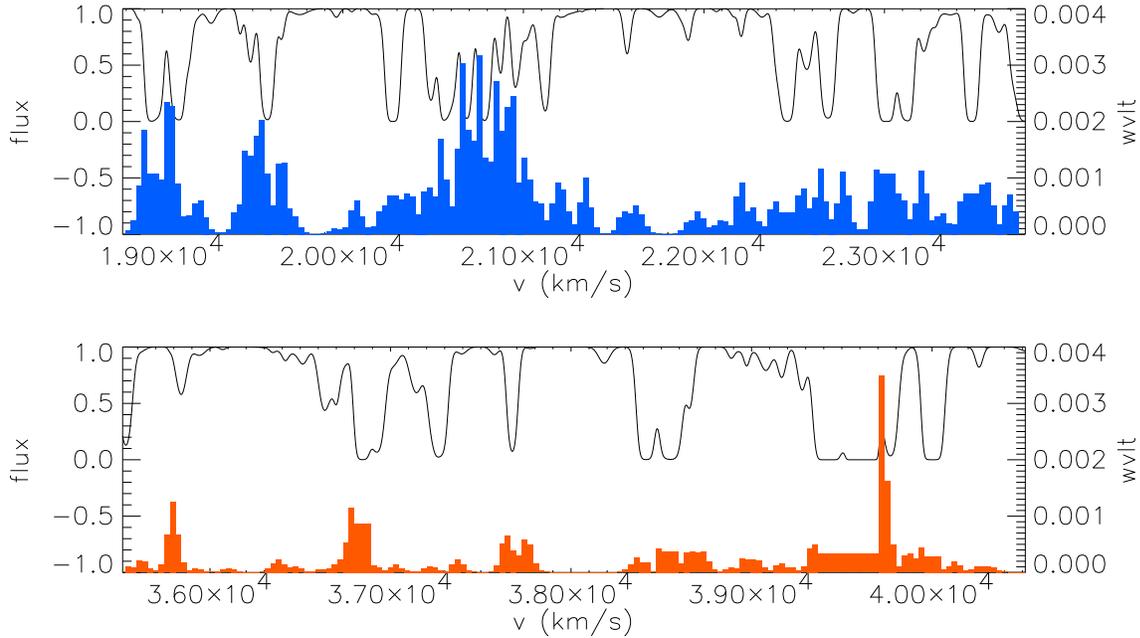}}
\end{picture}
\caption{Flux (lines, left hand scale) and wavelet amplitude averaged
over 100 km s$^{-1}$ (filled histogram, right hand scale) for a stretch
of spectrum of length 5000 km s$^{-1}$ from simulation C
($T_0=1.5\times 10^4$K, top panel) and the hotter simulation H
($T_0=2.2\times 10^4$K, bottom panel). The wavelet amplitudes tend to
be significantly larger in the colder model.}
\label{fig:fig1}
\end{figure*}

In summary: we start by fitting the original spectrum $F(\lambda)$ with
Voigt profiles, and reconstruct the spectrum from the fit, using only
those lines not identified as (or suspected to be) metal lines. For
this spectrum, $F_{\rm VP}(v)$, we compute the wavelet projection of its
normalised derivative, $F_{\rm VP}(\delta) = \sum_p
max_{(-1/2,0,1/2)\times\delta}( | \langle\psi(n,p)| \partial F_{\rm
VP}/\partial v\times (F_{\rm VP}+0.2)^{-1}(-\delta)\rangle
|\psi(n,p+\delta)\rangle|)$. We use the correlations in $F_{\rm
VP}(\delta)$ to quantify temperature variations along the
spectrum. Figure~\ref{fig:fig1} illustrates the procedure. It shows two
stretches taken from spectrum S1 of length 5000 km s$^{-1}$,
corresponding to model C and the 50 per cent hotter model H,
respectively.  The wavelet amplitudes tend to be significantly larger
in the spectrum taken from the colder model. The wavelet amplitudes can
therefore serve as a thermometer.

\subsection{Statistical analysis}
\subsubsection{Cumulative distribution of wavelet coefficients}
\begin{figure}
\setlength{\unitlength}{1cm} \centering
\begin{picture}(7,7)
\put(-1.5, -2.5){\includegraphics{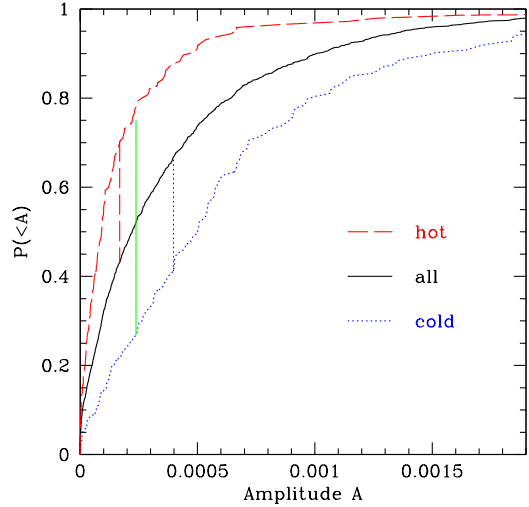}}
\end{picture}
\caption{Cumulative distribution of wavelet amplitudes for spectrum S1
(full line labelled all), its hot half (long dashed line) and its cold
half (dotted line). Vertical lines indicate the maximum differences
between the cumulative distributions.}
\label{fig:ksdist}
\end{figure}

\begin{figure}
\setlength{\unitlength}{1cm} \centering
\begin{picture}(7,4)
\put(-2., -3){\includegraphics{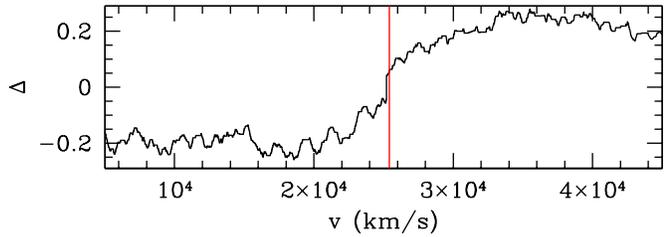}}
\end{picture}
\caption{Maximum difference $\Delta$ between the cumulative
distribution of $A$ in a window of size 5000 km s$^{-1}$ starting at
velocity $v$, and the cumulative distribution of $A$ over the whole
spectrum, for the simulated spectrum S1. Negative values of $\Delta$
indicate regions where the window has a larger fraction of high wavelet
amplitudes than the spectrum as a whole. $\Delta\sim -0.2$ for the
first half of the spectrum, and $\sim +0.2$b for the second half,
showing that $\Delta$ is able to recognise that the first half of
spectrum S1 is cold, and the second half is hot.}
\label{fig:ks_v}
\end{figure}

\begin{figure}
\setlength{\unitlength}{1cm} \centering
\begin{picture}(7,4)
\put(-2., -3){\includegraphics{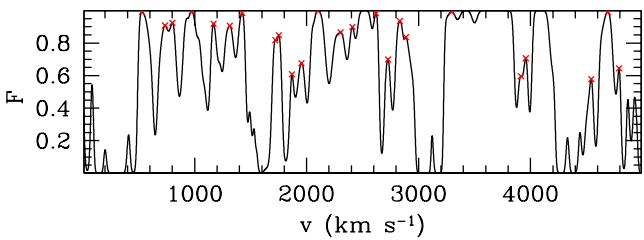}}
\end{picture}
\caption{Stretch of spectrum of QSO 1422+231. A piece of spectrum between
two crosses is identified as an \lq absorption feature\rq\ in the
randomisation procedure. Voigt profile fitted absorption lines with
central wavelengths that fall within the same absorption feature always
keep their relative spacing when randomising the spectrum, thereby
keeping the detailed line shape identical between original and
randomised spectrum.}
\label{fig:1422_ch}
\end{figure}

\begin{figure}
\setlength{\unitlength}{1cm} \centering
\begin{picture}(7,7)
\put(-1.5, -2.5){\includegraphics{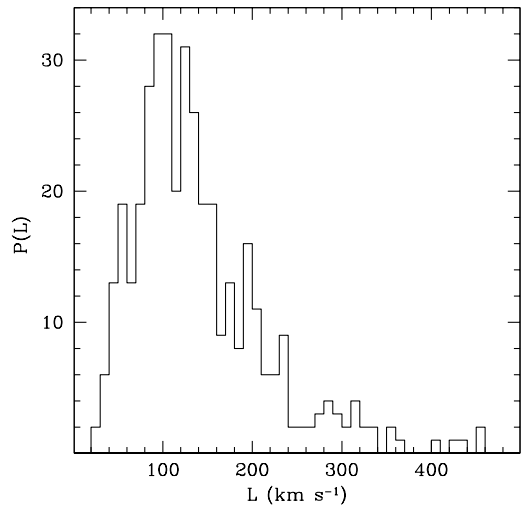}}
\end{picture}
\caption{Histogram of the lengths of absorption features identified in QSO
1422+231.}
\label{fig:chunck_hist}
\end{figure}

\begin{figure}
\setlength{\unitlength}{1cm} \centering
\begin{picture}(5,5)
\put(-1.5, -2.5){\includegraphics{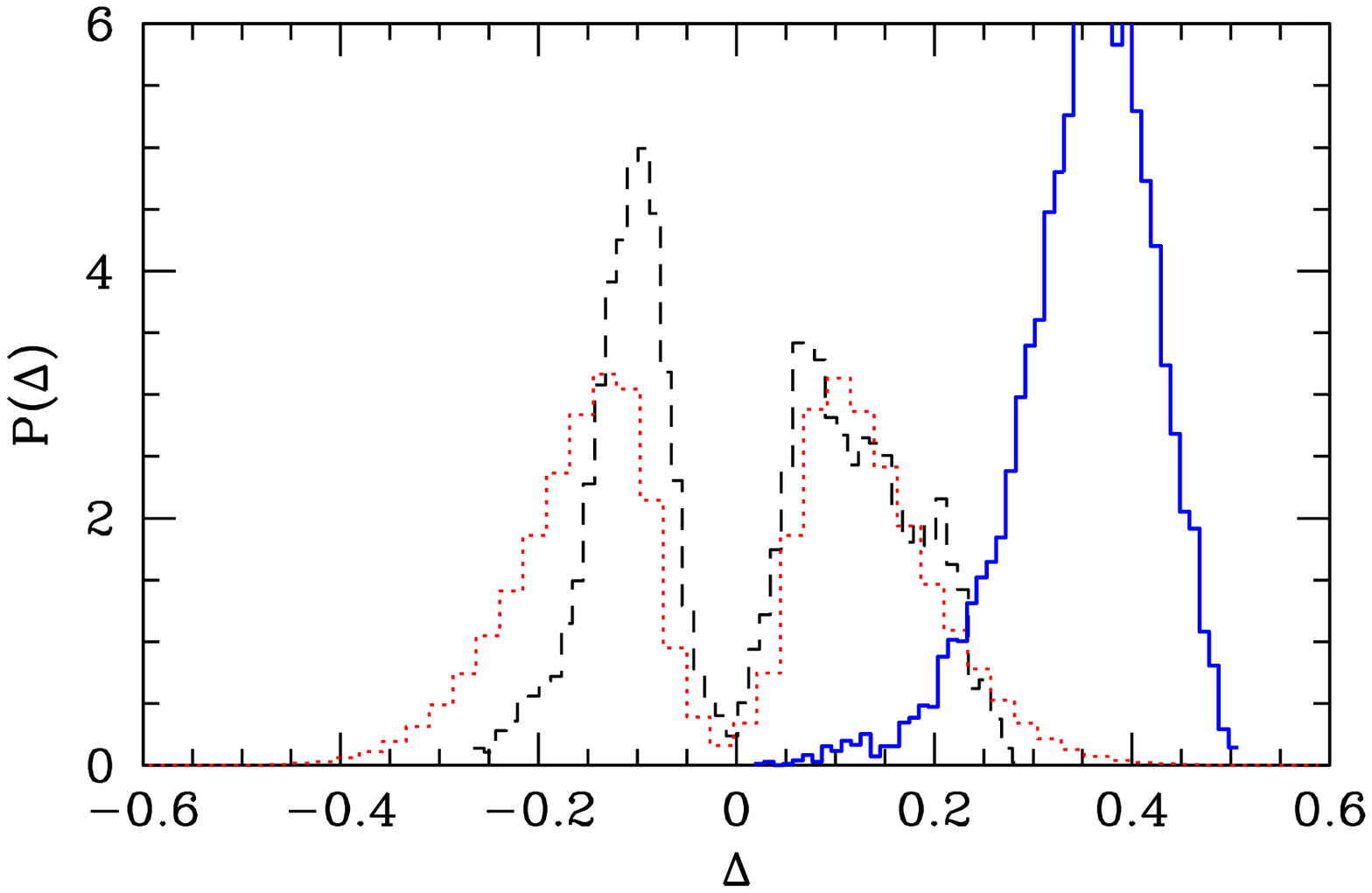}}
\end{picture}
\caption{Probability distribution $P(\Delta)$ for windows of size 5000
km s$^{-1}$ for windows drawn from 200 randomised spectra of spectrum
S1 (dotted line). Dashed line refers to $P(\Delta)$ for the original
spectrum S1, but only includes window pairs that fall in the same half
of S1, thick full line refers to window pairs that fall in different
halves. The dotted and dashed histograms are very similar, because both
refer to single temperature models. The histogram referring to
different temperature window pairs (full line) differs significantly
from the case where both windows refer to the same temperature. This
type of statistic is therefore useful in recognising regions with
different temperatures.}
\label{fig:delta_hist}
\end{figure}

In the previous section we illustrated the strong dependence of the
wavelet amplitude on the temperature of the gas --- cooler gas
generates a larger fraction of narrow lines with correspondingly higher
wavelet amplitudes $A$, on average. So we can now recognise temperature
fluctuations by identifying regions over which the wavelet amplitudes
are unusually large (a cold region), or unusually small (a hot
region). We will quantify the extent to which the wavelet amplitudes in
two regions of the spectrum differ, by $\Delta(v_1,\partial v_1;
v_2,\partial v_2)$, the maximum difference between the two cumulative
distributions of $A$ in the regions $[v_1,v_1+\partial v_1]$ and
$[v_2,v_2+\partial v_2]$ (velocities $v$ are in km s$^{-1}$). (Note
that $\Delta(v_1,\partial v_1; v_2,\partial v_2) =
-\Delta(v_2,\partial v_2; v_1,\partial v_1)$.) We will denote this
difference as $\Delta(v,\partial v)$ when one of the windows refers to
the whole spectrum.
	
The cumulative distribution $P(\le A)$ of the wavelet amplitude $A$ for
the two windows shown earlier in Fig.~\ref{fig:fig1} ($[1.9\times
10^4,5\times 10^3]$ and $[3.55\times 10^4,5\times 10^3]$ respectively)
is plotted in Fig.~\ref{fig:ksdist}, together with the cumulative
distribution of $A$ over the whole spectrum of S1.  The three
distributions appear to be quite different, as expected. The
differences $\Delta$ between the cumulative distributions are shown as
vertical lines. Note that Theuns \& Zaroubi (2000) also used $P(\le A)$
to characterise the $T-\rho$ relation.

The quantity $\Delta(v,\partial v)$ can be used as a (uncalibrated as
of yet) thermometer. In Fig.~\ref{fig:ks_v} we plot $\Delta(v,5\times
10^3)$ as a function of the starting position $v$ of the window for the
simulated spectrum S1. Notice that as long as $v$ is in the first half
of the spectrum -- in which case the window is drawn from the cold
simulation C -- $\Delta\sim -0.2$. There is a sudden transition around
$v\sim 2.5\times 10^4$ km s$^{-1}$ where $\Delta$ jumps to values $\sim
+0.2$, as the window starts to fall in the stretch of spectrum drawn
from the hotter model H. We want to stress again that, for a given
wavelet basis, $\Delta(v,5\times 10^3)$ follows uniquely from
projecting the spectrum on a set of basis functions.

Our aim is now to use $\Delta(v_1,\partial v_1;v_2,\partial v_2)$ to
find regions with different temperatures, and use $\Delta(v,\partial
v)$ as a measure of $T_0$. However, how can one judge whether a given
value of $\Delta$ is {\em statistically} significant? The usual way
(Kolmogorov-Smirnov test) to decide whether two data sets are drawn
from the same underlying distribution is to evaluate a particular
function of both $|\Delta|$, and of the effective number of degrees of
freedom $N$. Unfortunately, in the present case it is unclear how to
determine $N$, since pixels are correlated. One way to make progress is
to use randomised spectra, in which any temperature fluctuations have
been destroyed in the randomisation procedure, to {\em calibrate} the
distribution of $\Delta$ and hence determine how statistically
significant the difference seen in Fig.~\ref{fig:ksdist} is. Note that,
if a randomised spectrum, by chance, produces a a peculiar value of
$\Delta$, it is not due to temperature fluctuations, since the
temperture-density relation is not different from that of the rest of
the spectrum. The full procedure to assign a statistical significance
to unusual regions is described next.

\subsubsection{Randomised spectra}

The aim of this procedure is to produce new spectra from the data, in
which the absorption lines have the same shapes, but any {\em
correlation} between the lines is destroyed. One could in principle
randomise the positions of the Voigt profiles in Eq.~(\ref{eqn:vpfit}),
but the resulting spectra turn out to be quite different from the
original one. The reason is that many absorption features are composed
of several Voigt profiles (often with large error estimates for the
fitted parameters), some of which may be quite narrow and are
introduced to obtain a good fit. When randomising Voigt profiles, such
a narrow line will tend to occur on its own, and this causes clear
systematic differences between the randomised spectra and the original
one.

We have chosen instead to randomise the positions of the absorption
features (\lq absorption lines\rq\ as opposed to Voigt profiles, in
what follows) themselves. An absorption line is defined as a stretch of
spectrum between two local maxima in $F_{\rm VP}$. Voigt profiles are
now assigned to a line, if the centre of the profile falls between the
two maxima that define the line. Randomised spectra are then obtained
from randomising the positions of the {\em lines}, making sure that
lines do not overlap (but the Voigt profiles may), resampling them with
replacement. In the tests presented below, we will show that these
randomised spectra are not distinguishable from the original one, in
the {\em absence} of imposed temperature fluctuations.

An example of the identification of lines for part of the spectrum of
QSO~1422+231 is shown in Fig.~\ref{fig:1422_ch}, where local maxima are
indicated by crosses. We use only those local maxima where the flux
$F\ge 0.5$. A histogram of the widths of lines is shown for reference
in Fig.~\ref{fig:chunck_hist}. The typical widths of the lines is
$\sim 100$ km s$^{-1}$, with a tail to larger values.

Given these randomised spectra, we can compute the probability
distribution $P(\Delta)$, by randomly sampling $\sim 2\times 10^4$
window pairs for each spectrum. The result is plotted in
Fig.~\ref{fig:delta_hist} (dotted line). Superposed is $P(\Delta)$ for
the original spectrum S1, for window pairs drawn from the same half of
the spectrum (dashed line) and from different halves (full line). The
randomised spectra have a similar $P(\Delta)$ as the original spectrum
S1, when both windows are drawn from a single temperature region, but
the mixed temperature $P(\Delta)$ (full line) is very different. The
probability distribution $P(\Delta)$ from the randomised spectra allows
us to assign a statistical significance to a given value of $\Delta$.

\subsubsection{Cluster analysis}
\begin{figure}
\setlength{\unitlength}{1cm} \centering
\begin{picture}(8,8)
\put(-1.5, -0.5){\includegraphics{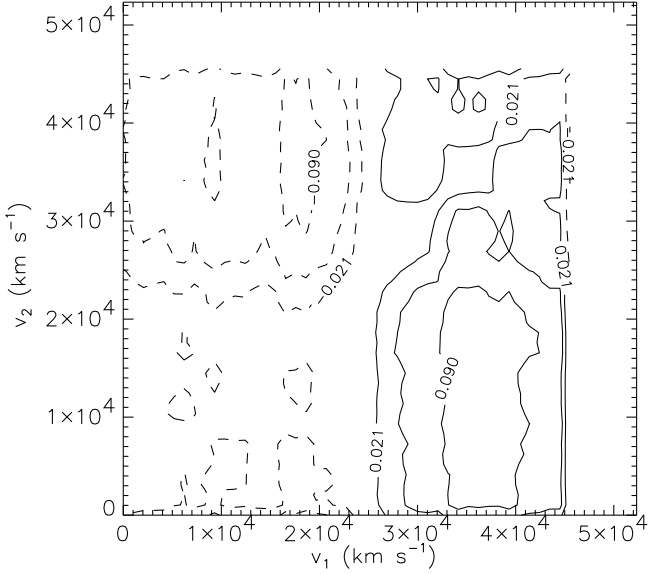}}
\end{picture}
\caption{Contour plot for the correlation field ${\cal Q}$ for spectrum S1.
Contours are drawn at levels 0.02, 0.05 and 0.09 for positive values
(full lines) and negative values (dashed lines). Window size $\partial
v=5000$ km s$^{-1}$. Contour level -0.021 delineates the cold region,
and level 0.021 the hot region.}
\label{fig:F}
\end{figure}

\begin{figure}
\setlength{\unitlength}{1cm} \centering
\begin{picture}(8,8)
\put(-1.5, -0.5){\includegraphics{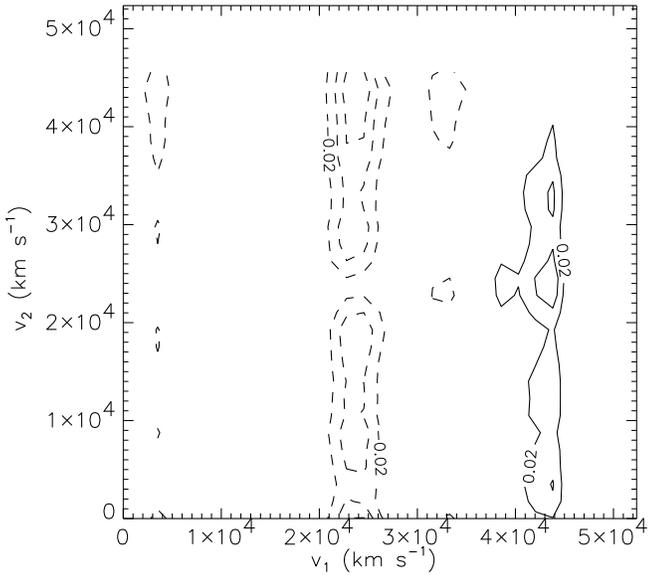}}
\end{picture}
\caption{Same as Fig.~\ref{fig:F} for spectrum S2. The cold gap in
the middle of the spectrum stands out.}
\label{fig:F_gap}
\end{figure}

In the previous sections we demonstrated how the difference $\Delta$
can be used as a thermometer, and how the statistics of $P(\Delta)$ can
be used to decide whether a difference in temperature between two
windows is statistically significant. Here we will combine these two
results to identify regions in the spectrum that have a statistically
significant different temperature as compared to the rest of the
spectrum.

In order to do so, we study correlations in the field
\begin{equation}
{\cal Q}(v_1,v_2;\partial v) = \Delta(v_1,\partial v) |\Delta(v_1,\partial v;v_2,\partial v)|\,.
\label{eqn:F}
\end{equation}
Recall that large values of $|\Delta(v_1,\partial v;v_2,\partial v)|$
suggest different temperatures between the two windows
$[v_1,v_1+\partial v]$ versus $[v_2,v_2+\partial v]$. Because we
multiply this quantity with $\Delta(v_1,\partial v)$ to obtain ${\cal Q}$, we
can also judge whether window $[v_1,v_1+\partial v]$ is cold or hot (in
which case ${\cal Q}\ll 0$ and ${\cal Q}\gg 0$, respectively), as we demonstrated in
Fig~.\ref{fig:ks_v}.

A contour representation of ${\cal Q}$ is plotted in Fig.~\ref{fig:F} for
spectrum S1, which shows that ${\cal Q}< 0$ in the top left hand corner, and
${\cal Q}> 0$ in the bottom right hand corner. Recalling our previous
results, this indicates that the first half of the spectrum is cold,
and has a significantly different temperature from the second half,
which consequently is hot. Recall that this is indeed the case for our
simulated spectrum S1.

The next step is to determine the extent of the region where the
temperature differs significantly from the rest of the spectrum. We do
this using a cluster analysis as follows.  Given the normalised
probability distribution $P(\Delta)$, determined from randomised
spectra without fluctuations, we determine $\Delta_f$ such that
$\int_{-\infty}^{\Delta_f} P(\Delta) d\Delta=f$, for $f=0.1$ and
$f=0.9$. That is, only 10 per cent of window pairs have $\Delta\le
\Delta_{0.1}$, and equally ten per cent of pairs have $\Delta\ge
\Delta_{0.9}$. We define ${\cal Q}_l\equiv |\Delta_{0.1}\,\Delta_{0.9}|$
as a good first indicator of whether a window has a peculiar
temperature, by comparing ${\cal Q}$ with ${\cal Q}_l$. Windows with
${\cal Q}< -{\cal Q}_l$ are likely to be peculiarly cold, those with
${\cal Q}> {\cal Q}_l$ peculiarly hot. In Fig.~\ref{fig:F}, the lowest
contour level shown is ${\cal Q}_l$, and indeed, these contour levels
delineate the low (cold) and high (hot) regions of the spectrum.

Another example of the field ${\cal Q}$, now for the spectrum S2, is
shown in Fig.~\ref{fig:F_gap}. Recall that S2 has a single cold region
located in $[2.2\times 10^4,2.7\times 10^4]$. This peculiar region
falls below the contour level $-{\cal Q}_l$ and is therefore neatly
detected by the procedure described so far. Note that there are other,
smaller, regions where $|{\cal Q}|\ge {\cal Q}_l$ in this example.

We now define a {\em cluster} as a connected region where $|{\cal
Q}|\ge {\cal Q}_l$. In practise, we identify clusters by interpolating
${\cal Q}$ onto a grid of grid size $\partial v/5$ using
cloud-in-cell interpolation. A cluster is then a set of connected grid
cells\footnote{Grid cells that neighbour each other either horizontally
or vertically, are called connected.}, which each have $|{\cal Q}|\ge
{\cal Q}_l$. The {\em weight} of a cluster is the integral of ${\cal
Q}$ over its cells. The motivation for doing this, is that a region of
unusual temperature will have a large extent in a plot such as
Fig.~\ref{fig:F}, and so will have a large weight. We will use the
randomised spectra to judge the statistical significantly of those
weights.

When identifying clusters, we impose \lq periodic boundary
conditions\rq~ in the vertical direction (i.e., cells at the upper and
lower extreme, but the same horizontal position $v_1$, are also
neighbouring cells). This may seem strange at first, but consider the
cluster identified in Fig.~\ref{fig:F_gap}. Without periodic boundary
conditions, there would be two clusters (of roughly half the size)
above the cold region $[2.2\times 10^4,2.7\times 10^4]$. The reason
that these two clusters are not connected, but appear to have a hole
around $v_2\sim 2.5\times 10^4$ is of course because there, $v_1\sim
v_2$, and both windows refer to the {\em same} stretch of spectrum --
hence ${\cal Q}\sim 0$. So clearly that region of the plane {\em is
not} peculiar as it does not refer to windows of different
temperatures. However, if the cold region had been closer to the start,
or the end of the spectrum, there wouldn't have been a hole, and hence
the cluster had been identified with nearly twice the weight it has
now. This is clearly not what we want. Imposing periodic boundary
conditions connects these two clusters again into one. In this case,
the weight of the cluster does {\em not} depend on its position. Note
that clusters that correspond to low temperatures have a negative
weight.

After cluster identification, we can introduce the new field, ${\cal
C}$, defined on the grid, where ${\cal C}(v_1,v_2)=0$ if the cell does
not fall into a cluster, and equals the weight of the cluster, if it
does. We finally project ${\cal C}$ onto the spectrum,
\begin{equation}
C_P(v_1) = \sum_{v_2} {\cal C}(v_1,v_2)\,.
\label{eqn:C_P}
\end{equation}
For a given velocity, $C_P(v)$ is a measure of the fraction of spectrum
that has a significantly different temperature. The level of
significance of a given value $C_P(v)$ is obtained by performing the
same analysis on all the randomised spectra, from which we can compute
the probability distribution $P(C_P)$. This concludes the description
of the statistical analysis.

\subsection{Temperature calibration}
\begin{figure*}
\setlength{\unitlength}{1cm} \centering
\begin{picture}(12,6.6)
\put(-3, -6.0){\includegraphics{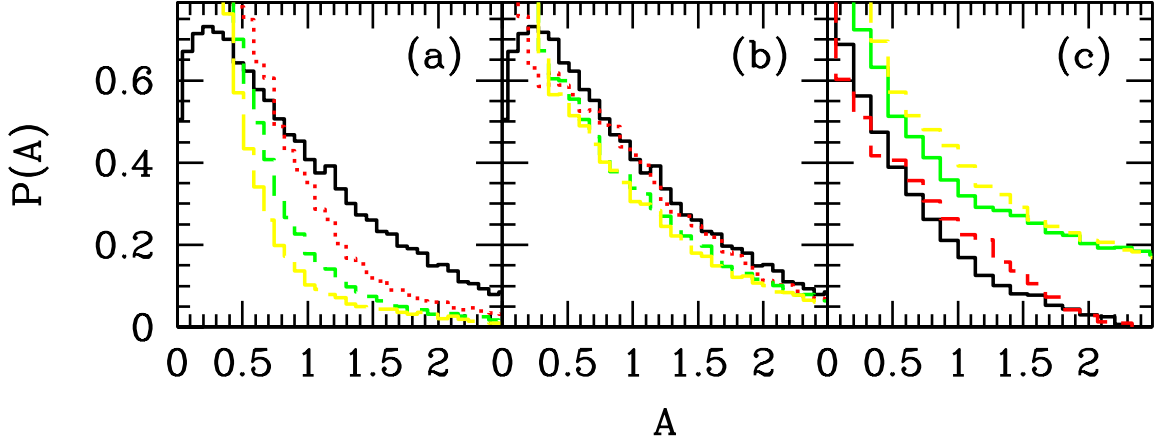}}
\end{picture}
\caption{Dependence of the probability distribution $P(A)$ of wavelet
amplitudes on $T_0$ and $\gamma$. Panel (a): $P(A/\langle A_1\rangle)$
versus $A/\langle A_1\rangle$ for models with ${\rm log}_{10}
T_0=[4.0,4.1,4.2,4.3]$ (full, dotted, short-dashed and long-dashed
lines respectively) and $\gamma=1.4$. The amplitudes have been scaled
by the mean wavelet amplitude $\langle A_1\rangle$ of the coldest
model. In panel (b), the distribution $P(A/\langle A\rangle)$ is scaled
using the mean of the distribution. In panel (c) we show the normalised
distributions $P(A/\langle A\rangle)$ for models with $({\rm log}_{10}
T_0,\gamma)=(4.18,1)$ and (4.18,5/3) (lower curves, full and dashed
line respectively, off-set vertically by -0.1) and $({\rm log}_{10}
T_0,\gamma)=(4.34,1)$ and (4.34,5/3) (full and dashed lines
respectively, off-set vertically by 0.1). Panel (a) shows that $P(A)$
depends on $T_0$, with hotter models having a smaller fraction of large
wavelet amplitudes, as expected. Panel (b) demonstrates that the
difference in shape is mostly due to the difference in mean $\langle
A\rangle$ of the distribution. Finally, panel (c) shows that $\gamma$
has an effect on the shape of $P(A/\langle A\rangle)$ as well, such
that models with a larger $\gamma$ are larger around the mean of the
distribution.}
\label{fig:adist}
\end{figure*}

\begin{figure}
\setlength{\unitlength}{1cm} \centering
\begin{picture}(7,7)
\put(-1.5, -2.5){\includegraphics{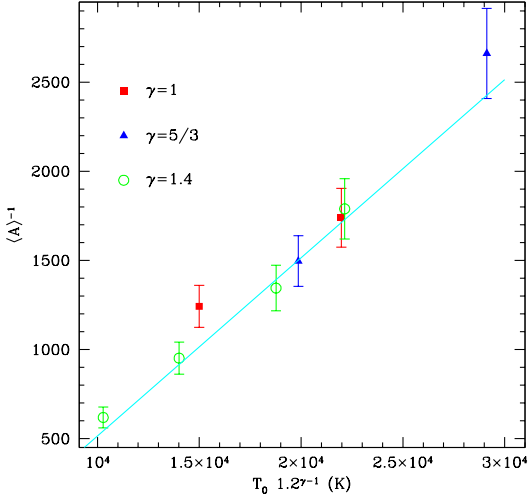}}
\end{picture}
\caption{Temperature $T_{1.2}\equiv T_0\times 1.2^{\gamma-1}$ at an
overdensity 1.2, versus the inverse $\langle A\rangle^{-1}$ of the
wavelet amplitude for models with different values of $T_0$ and
$\gamma$ at a redshift $z=3$. The values of $\gamma$ are indicated in
the panel. The error bars assume 30 per cent uncertainty on $\langle
A\rangle$ per 5000 km s$^{-1}$. The line is a least squares fit to the
points, demonstrating the scaling $T_{1.2}\propto \langle
A\rangle^{-1}$.}
\label{fig:aversust}
\end{figure}

The method described so far identifies regions of significantly
different temperatures. We can get an idea of how much the temperatures
differ by examining in more detail the probability distribution $P(A)$
of wavelet amplitudes. The shape of this distribution depends mainly on
its average $\langle A\rangle$, as pointed out by Zaldarriaga (2001),
and demonstrated in Fig.~\ref{fig:adist}. There is an additional
dependence on the slope $\gamma$ of the $T-\rho$ relation, in that
models with larger $\gamma$ have a higher fraction of wavelets around
the mean.

In turn, $\langle A\rangle$ correlates strongly with the temperature
$T_\delta\equiv T \delta^{\gamma-1}$ of the gas, at some overdensity
$\delta\equiv\rho/\langle\rho\rangle$. For the mean absorption
appropriate for $z=3$, $\delta\approx 1.2$. Figure~\ref{fig:aversust}
shows that $T_\delta\propto \langle A\rangle^{-1}$, for simulated
models with a range of values for $T_0$ and $\gamma$. So without
calibrating the $\langle A\rangle-T$ relation with simulations, one can
still estimate temperature ratios. If one is willing to normalise the
relation using simulations, then it is possible to measure $T_{1.2}$
directly.

In the linear regime, the baryonic overdensity $\delta_{\rm IGM}$ is
smoothed with respect to the dark matter field $\delta_{\rm DM}$
according to $\delta_{\rm IGM}(k)\propto \delta_{\rm DM}(k)/k^2T$, at
high wavenumbers $k$ (e.g. Bi and Davidsen 1997). The wavelet amplitude
$\langle A\rangle$ is a measure of the r.m.s. of $\delta_{\rm IGM}$ on
small scales, which explains the scaling $\langle A\rangle\propto
\langle\delta_{\rm IGM}^2\rangle^{1/2}\propto T^{-1}$.

The mean $\langle A\rangle$ can be directly computed for a given
spectrum, and the linear relation of Fig.~\ref{fig:aversust} can then
be used to estimate $T_{1.2}$. In the simulations we measured a
variation of $\sim 30$ per cent in $\langle A\rangle$ over regions of
$\partial v=5000$ km s$^{-1}$, in single temperature models. The
corresponding error bars are indicated in Fig.~\ref{fig:aversust}. When
applying this linear scaling to data of velocity extent $\partial v$,
we will also assume a similar error $\sim 30 (\partial v/5000 {\rm km
s}^{-1})^{-1/2}$ per cent. The values $\gamma=1$ and $\gamma=5/3$ are
likely to span the range encountered when applying this calibration to
real data.

\subsubsection{Summary and illustration}
\begin{figure}
\setlength{\unitlength}{1cm} \centering
\begin{picture}(7,9)
\put(-1.5, -2){\includegraphics{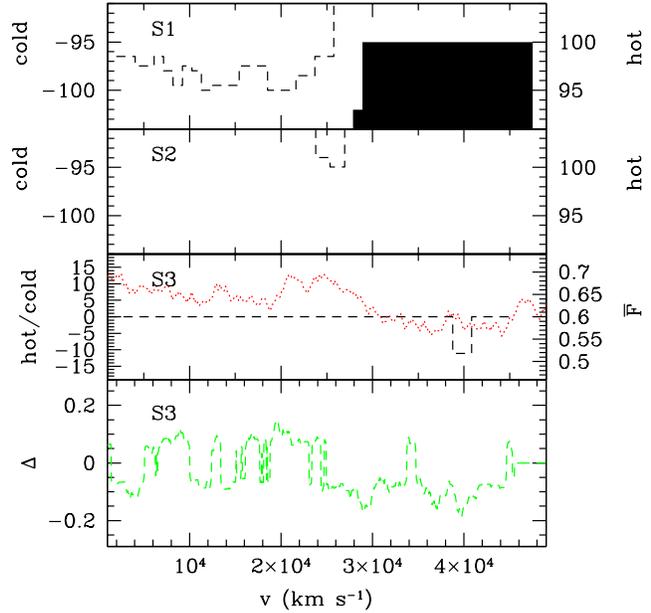}}
\end{picture}
\caption{Statistical significance that the temperature in a window of
size 5000 km s$^{-1}$ starting at velocity $v$ differs from the
spectrum as a whole. Outlined histogram (left hand scale) is
significance in per cent for cold regions, filled histogram (right hand
scale) for hot regions. Dotted line (left hand scale) denotes the flux,
smoothed over 5000 km s$^{-1}$. Both the hot and the cold region in S1
(top panel) are assigned a significance of 100 per cent. The cold gap
in spectrum S2 has a significance of 95 per cent. Finally, no
significant region is detected in the single temperature spectrum S3
(bottom panel). The bottom panel shows the $\Delta$ statistic for S3,
which can be compared with the corresponding curve for S1 in
Fig.~\ref{fig:ks_v}}
\label{fig:Hep1_Hep3_wvlt}
\end{figure}

The method we have devised consists of three steps.

\noindent (1) Compute the contribution of a narrow wavelet to the
derivative of the spectrum. We used simulation to show that the
amplitude $A$ of such a wavelet anti-correlates strongly with the
temperature of absorbing gas (Fig.~\ref{fig:fig1}). The cumulative
distribution $P(<A)$ can be used to characterise the $\rho-T$ relation.
For example, we used the maximum distance $\Delta$ between the
cumulative distributions of two stretches of spectrum to judge which
stretch corresponds to higher temperatures (Fig.~\ref{fig:ks_v}).

\noindent (2) Generate random spectra from the original data by
scrambling the line list. These random spectra are used to compute the
probability distribution $P(\Delta)$ for spectra {\em without}
temperature fluctuations.

\noindent (3) Identify unusual regions in the spectrum, that correspond
to large values of $|{\cal Q}|$, where ${\cal Q}(v_1,v_2,\partial v)$ is defined in
Eq.~(\ref{eqn:F}). Use a clustering analysis to better characterise the
size of these regions. Project the resulting clusters on to the
spectrum to obtain $C_P(v)$ (Eq.~\ref{eqn:C_P}) and use the randomised
spectra to associate the statistical significance.

An estimate of the temperature ratio of different regions can be
obtained by comparing the mean wavelet amplitudes and using the
relation $T\propto \langle A\rangle^{-1}$.

The final result of applying the above procedure to the simulated
spectra S1, S2 and S3, is shown in Fig.~\ref{fig:Hep1_Hep3_wvlt}. For
each value of $C_P(v)$, we find the fraction $f$ of random spectra that
have a comparable cluster, and plot the significance $100(1.-f)$ in per
cent (or $100(f-1.)$ for cold regions), both for hot regions (filled
histogram, right hand scale) and cold regions (histogram, left hand
scale). The hot and cold regions in spectrum S1 both have a
significance of $\sim 100$ per cent, meaning that none of the (200 in
this case) randomised spectra contain clusters that large. The cold gap
in spectrum S2 is significant at the 95 per cent level.

The different $\rho-T$ relations in the spectra S1 and S2 were imposed
on the simulations in post processing. This is likely to underestimate
the differences in line widths, compared to models in which the
temperature is different {\em during} the simulation as well. For
example, Theuns et al. (2000) showed that line widths for simulation
with identical {\em imposed} $\rho-T$ relations are measurably
different if the underlying simulations had different temperatures. The
reason is that simulations with different temperatures vary in the
amount of Jeans smoothing and pressure induced peculiar velocities,
which also contribute to the line widths. These effects are not
captured by changing the $\rho-T$ relation, which only influences the
thermal broadening.

Finally, spectrum S3 drawn from a single temperature model does not
contain any regions significant to more than 10 per cent. Recall that
this simulated spectrum has a large jump in optical depth, with mean
fluxes $\bar F\sim 0.64$ and $\sim 0.60$ for the first, respectively
second half of the spectrum. This latter test demonstrates that the
procedure of randomising spectra works well --- the original spectrum
is equivalent to the randomised spectra for the statistic we
investigate here. Note also that the method is not very sensitive to
the mean effective optical depth. We apply the analysis procedure to
data in the next Section.

\section{Thermal state of the IGM}
\subsection{Redshift range $z=[3,3.6]$}
\label{sect:highz}
\begin{figure}
\setlength{\unitlength}{1cm} \centering
\begin{picture}(7,5)
\put(-2., -3){\includegraphics{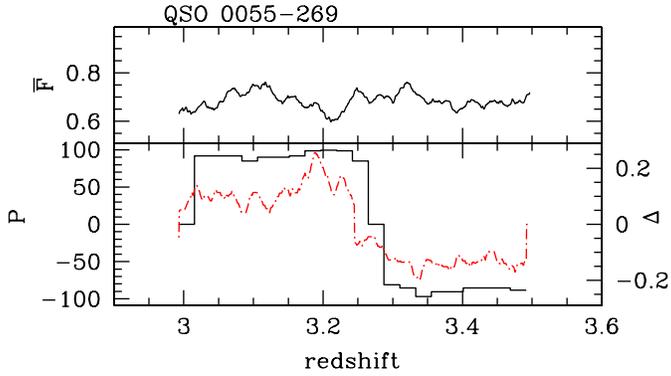}}
\end{picture}
\caption{Wavelet analysis of the spectrum of QSO 0055--269. Top panel:
mean flux averaged over a window of $\partial v=8000$ km
s$^{-1}$. Bottom panel: temperature measure $\Delta(v,\partial v)$
(dot-dashed line, right hand scale), and statistical significance
$P(C_P)$, where $C_P$ is defined in Eq.~(\ref{eqn:C_P}), of temperature
fluctuations in per cent (histogram, left hand scale). The low redshift
half of the spectrum is unusually hot at the 99.5 per cent level, and
the high redshift half of the spectrum is unusually cold at the 97 per
cent level. The jump in temperature appears to be very sudden and
occurs at a redshift $z\sim 3.3$. The mean absorption $\bar F$ does not
appear to undergo a similar strong evolution.}
\label{fig:0055}
\end{figure}

\begin{figure}
\setlength{\unitlength}{1cm} \centering
\begin{picture}(7,5)
\put(-2., -3){\includegraphics{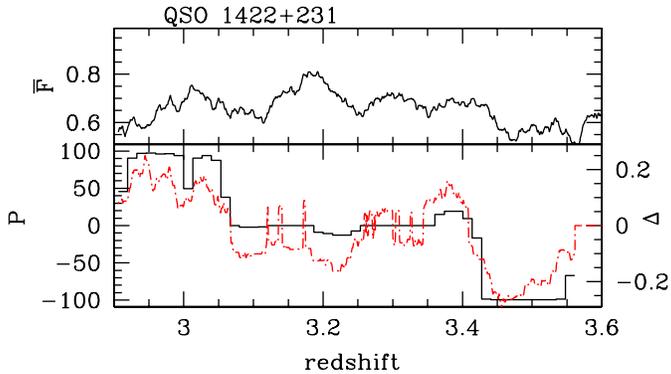}}
\end{picture}
\caption{Same as Fig.~\ref{fig:0055} for QSO 1422+231, but for a window
size of $\partial v=5000$ km s$^{-1}$. Below $z\sim 3.05$, the spectrum
is unusually hot at the 97.5 per cent level. Above $z\sim 3.4$ it is
unusually cold at the 99.5 per cent level. In between, there appear to
be large fluctuations in $\Delta(v,\partial v)$, which however are not
statistically significant, when compared to the spectrum as a whole.}
\label{fig:1422}
\end{figure}

\begin{figure}
\setlength{\unitlength}{1cm} \centering
\begin{picture}(7,5)
\put(-2., -3){\includegraphics{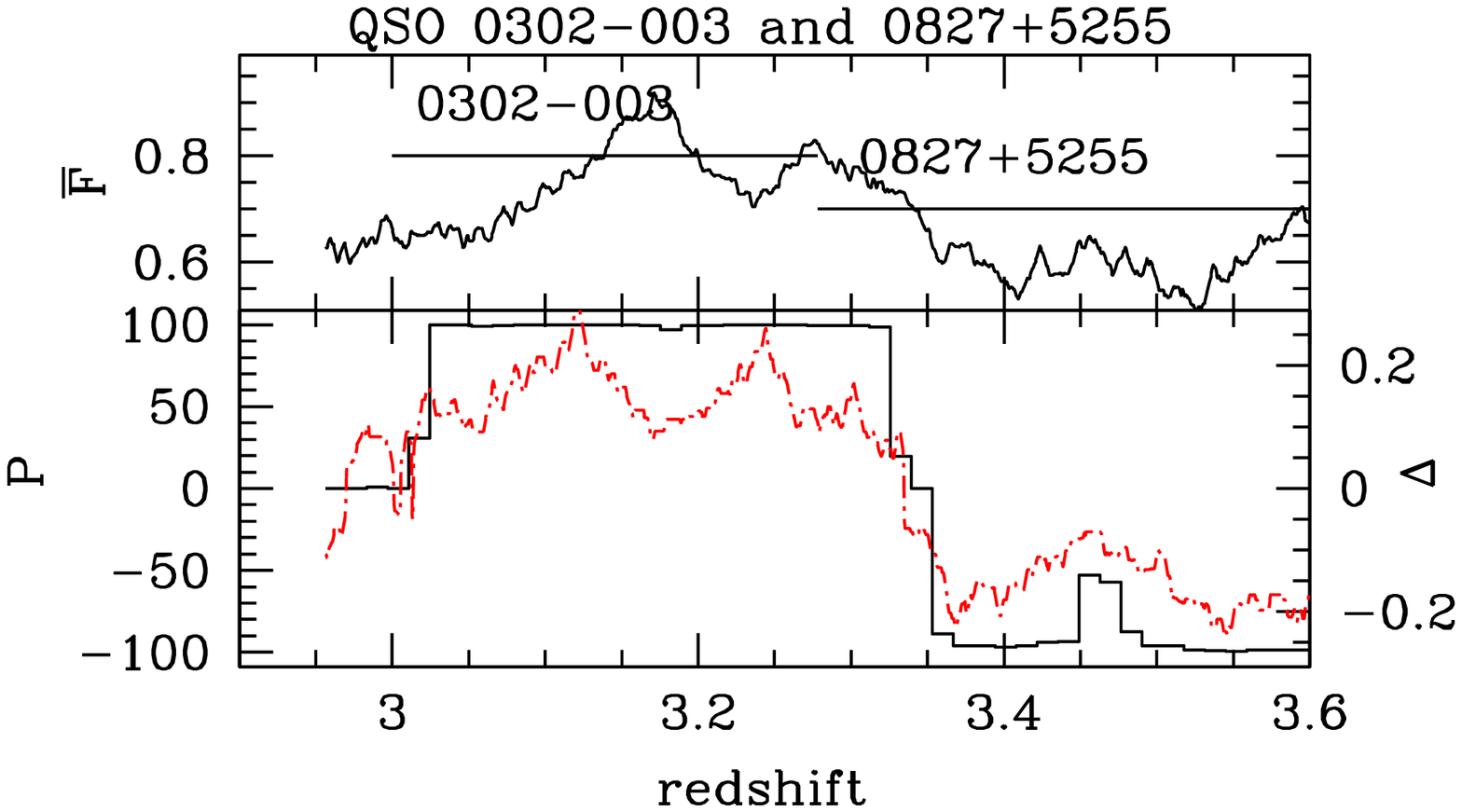}}
\end{picture}
\caption{Same as Fig.~\ref{fig:1422} for QSO 0302-003 (interval
$z=[3,3.27]$ and QSO APM~0827+5255 (interval $z=[3.27,3.7]$). The spectrum below
$z\sim 3.35$ is hot at the 100 per cent level, and above $z\sim 3.35$
it is cold at the 99.5 per cent level.}
\label{fig:0302+0827}
\end{figure}

\begin{figure}
\setlength{\unitlength}{1cm} \centering
\begin{picture}(7,5)
\put(-2., -3){\includegraphics{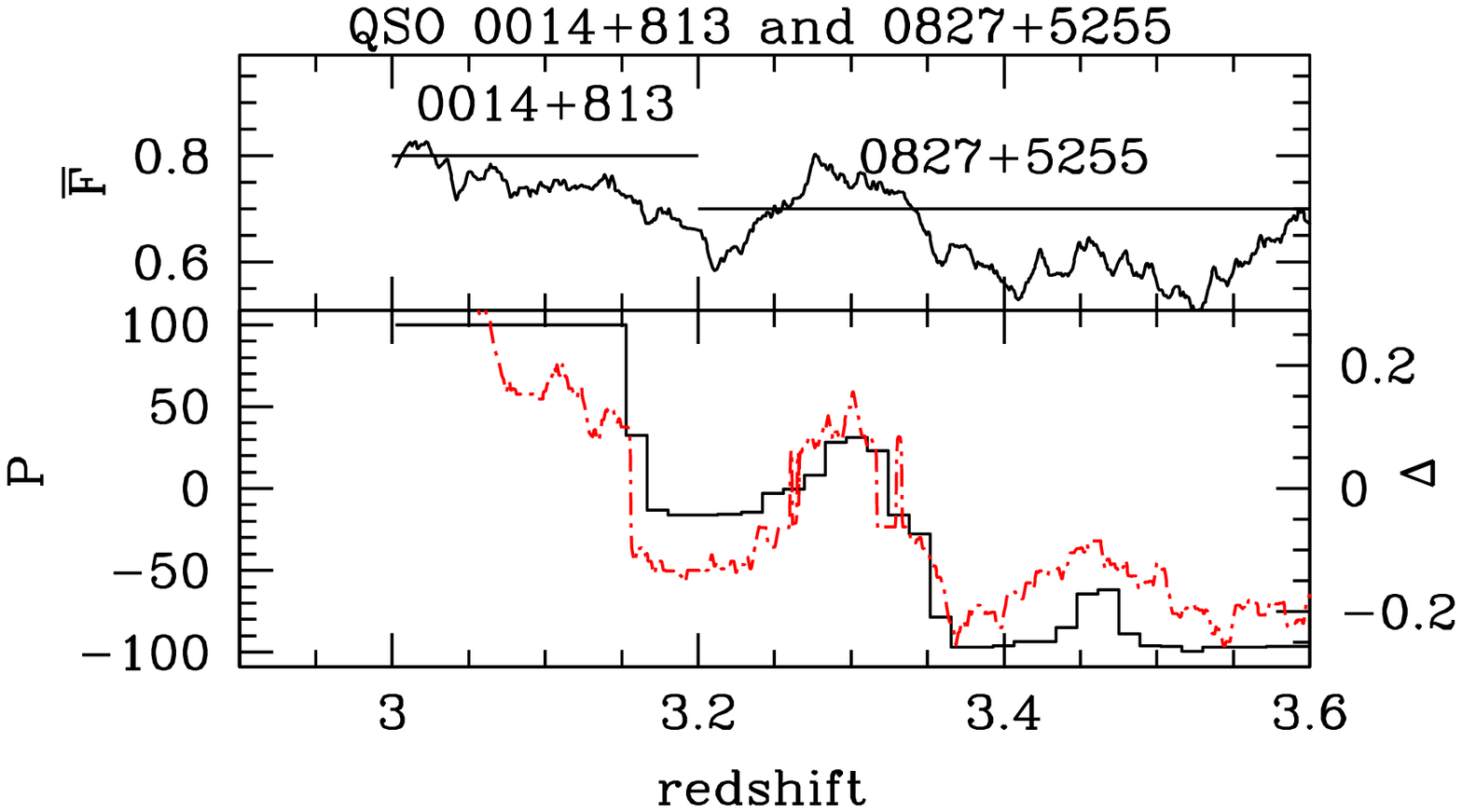}}
\end{picture}
\caption{Same as Fig.~\ref{fig:0302+0827} but for QSO 0014+813 (interval
$z=[3,3.2]$ and QSO APM~0827+5255 (interval $z=[3.2,3.7]$). The spectrum below
$z\sim 3.15$ is hot at the 100 per cent level, and above $z\sim 3.35$
it is cold at the 99.5 per cent level.}
\label{fig:0014+0827}
\end{figure}

\begin{figure}
\setlength{\unitlength}{1cm} \centering
\begin{picture}(7,7)
\put(-2., -4.){\includegraphics{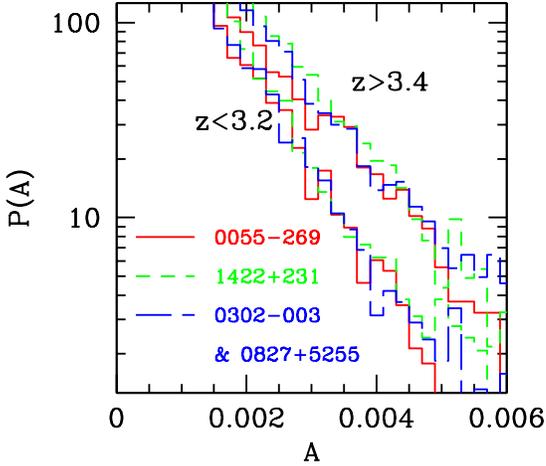}}
\end{picture}
\caption{Probability distribution of wavelet amplitudes, $P(A)$, for
the combined spectrum of QSOs 0302-003 and APM~0827+5255, and for QSOs
1422+231 and 0055-269. The high redshift halves of the spectra
($[3.37,3.6]$,$[3.45,3.5]$ and $[3.35,3.5]$ respectively) and the low
redshift halves ($[3.05,3.3]$,$[2.95,3.05]$,$[3.05,3.25]$) are shown
separately. $P(A)$ for spectra at the same redshift are very similar,
but there is a difference between the probability distributions at
different redshifts.}
\label{fig:wavdist}
\end{figure}

\begin{figure}
\setlength{\unitlength}{1cm} \centering
\begin{picture}(7,15)
\put(-2., -6.){\includegraphics{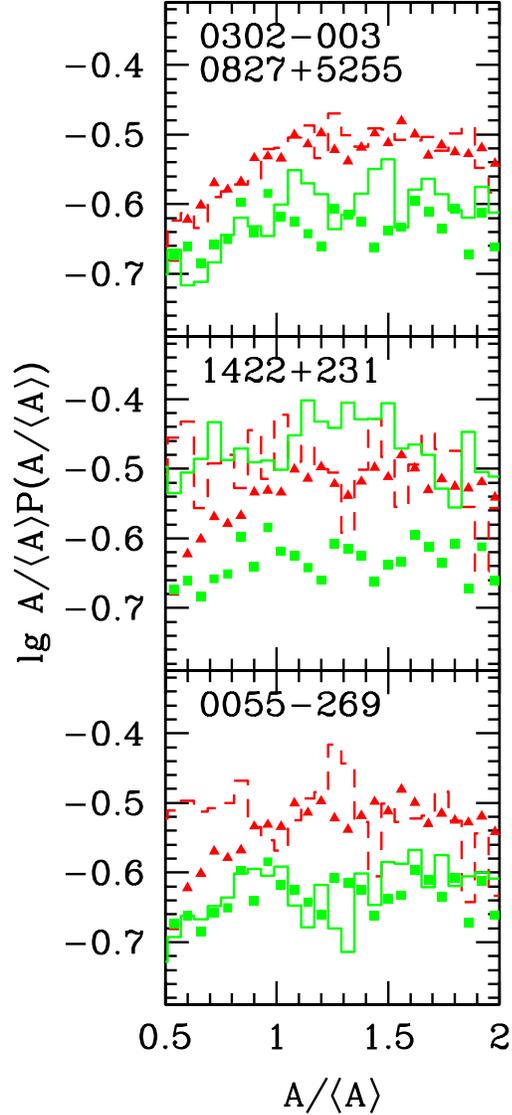}}
\end{picture}
\caption{Same as Fig.~\ref{fig:wavdist}, but for the distribution of
wavelet amplitudes in units of the mean amplitude, $A/\langle A\rangle
P (A/\langle A\rangle)$. Dashed lines refer to the high redshift parts of
the spectrum, full lines to the low redshift halves. Filled triangles and
filled squares refer to simulations with $(log T_0,\gamma)=(4.18,5/3)$
and $(4.34,1)$ respectively. The high redshift halves are well
represented by the colder model with a steep $T-\rho$ relation, the
lower redshift halves are better fit by the hotter, isothermal model.}
\label{fig:wavdes}
\end{figure}

\begin{figure}
\setlength{\unitlength}{1cm} \centering
\begin{picture}(8,9)
\put(-2., -3.){\includegraphics{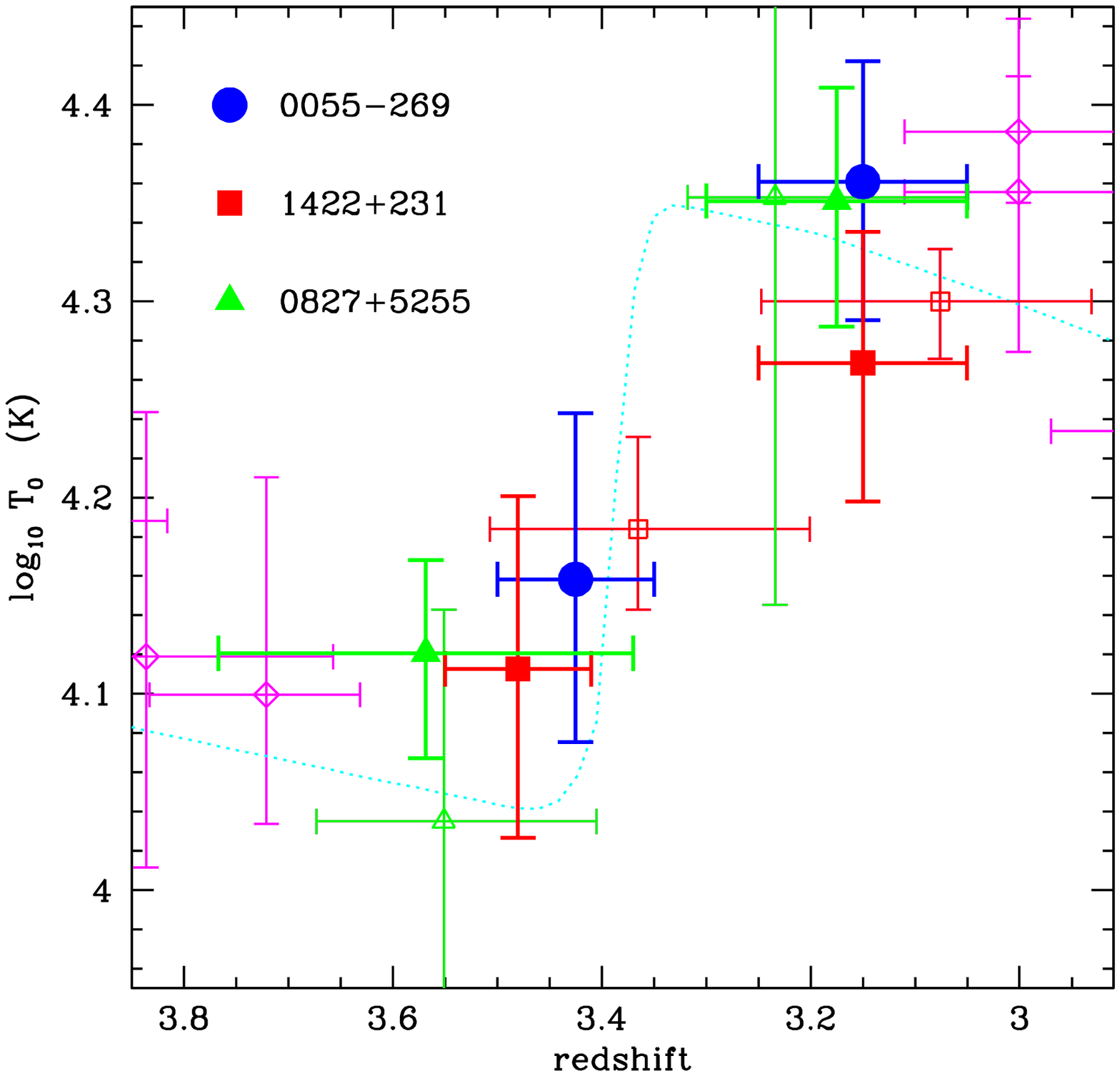}}
\end{picture}
\caption{Temperature at the mean density, $T_0$, versus redshift. Open
symbols with error bars are taken from Schaye et al. (2000). Filled
symbols with error bars and thicker lines refer to estimates of $T_0$
based on the mean wavelet amplitudes and the calibration of
Fig.~\ref{fig:aversust}, with errors on $T_0$ of 30 per cent for a
stretch of spectrum of 5000 km s$^{-1}$. Errors in the redshift
direction refer to the redshift extent of the sampled region. Symbol
types refer to the QSOs (note that APM~0827+5255 and 1422+231 were used in
the analysis of Schaye et al. 2000 as well.) We have assumed $\gamma=1$
below $z=3.3$, and $\gamma=5/3$ above $z=3.3$ to convert $T_{1.2}$ to
$T_0$. Dotted line is the temperature evolution in a simulation where
\Hep reionizes at redshift $\sim 3.4$. It fits the data very well.}
\label{fig:tempnorm}
\end{figure}

Two of our QSO spectra span the region around $z\sim 3.2$ where Schaye
et al. (2000) claimed an increase in $T_0$. We have prepended a stretch
of spectrum from QSO 0302-003 and of 0014+813 to that of APM~0827+5255, to
have two more spectra that straddle that redshift $z\sim 3.2$. We use
these spectra to investigate whether our new method independently
suggests a change in $T_0$ around redshift $z\sim 3.2$. For each of the
spectra, we have generated 200 randomised spectra to judge the
significance of the temperature fluctuations as described earlier.

QSO 0055--269 does indeed appear to show a very sudden increase in
$T_0$ around $z\sim 3.3$ (Fig.~\ref{fig:0055}). Its low redshift half
is unusually hot at the 99.5 per cent level, and the high redshift half
is unusually cold at the 97 per cent level. It is worth noting that the
low redshift half contain two voids at $z\sim 3.11$ and 3.28 of sizes
$\sim 17$ and 15 $h^{-1}$ Mpc respectively (deceleration parameter
$q_0=0.1$, Kim et al 2001b). This sudden transition looks very similar
to what we had for our simulated spectrum S1 (but there the low
redshift half of S1 was cold). The identification of a statistically
significant change in $T_0$ does not require the use of
simulations. However, once we need to decide {\em by how much} $T_0$
changes, simulations are needed in order to calibrate the relation
between the wavelet distribution and $(T_0,\gamma)$.

QSO 1422+231 also shows an increase in $T_0$, albeit not so suddenly
(Fig.~\ref{fig:1422}). As for QSO 0055=269, the top (bottom) half is
unusually cold (hot), at the 99.5 (97.5) per cent level. Note that
there are large fluctuations in $\Delta$ in between the hot and cold
regions, in the interval $z=[3.1,3.4]$. In particular, it would be
tempting to identify the hot region around $z\sim 3.37$ as being heated
early, and/or the cold region around $z\sim 3.2$ as remaining cold
slightly longer. However, the regions around $z\sim 3$ and $z\sim 3.5$
are much larger, hence far more significant. When randomised spectra
are generated from this spectrum, many will actually contain large
cold/hot regions as well, making the apparent fluctuations around
$z\sim 3.2$ not statistically significant.

The combined spectrum of QSOs 0302--003 (from $z=[3,3.27]$) and
APM~0827+5255 (from $z=[3.27,3.7]$) is shown in
Fig.~\ref{fig:0302+0827}. There is a sudden change in temperature at
$z=3.3$, with a significance of 100 per cent for the low redshift hot
half, and 99.5 per cent for the high redshift colder region. QSO
0302--003 contains a well known void at $z\sim 3.17$ (Dobrzycki \&
Bechtold 1991) which is visible in Fig.~\ref{fig:0302+0827} as a region
of low absorption. We note that the higher resolution data of Kim et al
(2001b) suggest a smaller size for the void than the earlier data. In
any case, comparing the mean absorption with the temperature measure
$\Delta$, it is clear that the high inferred temperatures do {\em not}
just result from the presence of these voids, although they might
contribute to it.

The combined spectrum of QSOs 0014+813 (from $z=[3,3.2]$) and APM~0827+5255
(from $z=[3.2,3.7]$) is shown in Fig.~\ref{fig:0014+0827}. Below
$z=3.15$, this spectrum is hot at the 100 per cent level, above
$z=3.35$ it is cold at the 99.5 per cent level. In between, there are
large scale fluctuations similar to what we obtained for QSO 1422+231.

The probability of wavelet amplitudes, $P(A)$, is plotted in
Fig.~\ref{fig:wavdist} for the unusual regions identified in these
spectra. Consistent with the above evidence of a temperature jump, we
find a change in shape of $P(A)$ around $z\sim 3.3$. Comparing the
normalised distribution, $P(A/\langle A\rangle)$, with the simulations,
we find good agreement between the high redshift halves and the
simulation with $({\rm log} T_0,\gamma)=(4.18,5/3)$, and the low
redshift half and the isothermal model $({\rm log}
T_0,\gamma)=(4.34,1)$ respectively (Fig.~\ref{fig:wavdes}). Using these values for $\gamma$, we
can apply the calibration between $\langle A\rangle$ and $T_{1.2}\equiv
T_0 1.2^{\gamma-1}$ from Fig.\ref{fig:aversust} to obtain an estimate
for $T_0$. The result is in excellent agreement with the temperature
evolution determined by Schaye et al. (2000) using the $b-N$ cut-off
(Fig.~\ref{fig:tempnorm}). Above redshift $z\sim 3.4$, the temperature
$T_0\sim 10^{4.1\pm 0.15}$, and below that $T_0\sim 10^{4.3\pm 0.15}$,
an increase of around 60 per cent.

We conclude that all four spectra show evidence for a change in
temperature, significant at the $\sim 99$ per cent level, over a
relatively narrow redshift interval around $z\sim 3.3\pm 0.2$. The
temperature increase is 60 $\pm$ 14 per cent. Such an evolution is
clearly not consistent with photo-heating of a highly ionised,
expanding medium, in which case the temperature would {\em smoothly
decrease} towards lower $z$, but is consistent with a sudden entropy
injection in the IGM resulting from \Hep reionization.

\subsection{Redshift range $z=[2.6,3.2]$}
\label{sect:intz}
\begin{figure}
\setlength{\unitlength}{1cm} \centering
\begin{picture}(7,7)
\put(-2.5, -3){\includegraphics{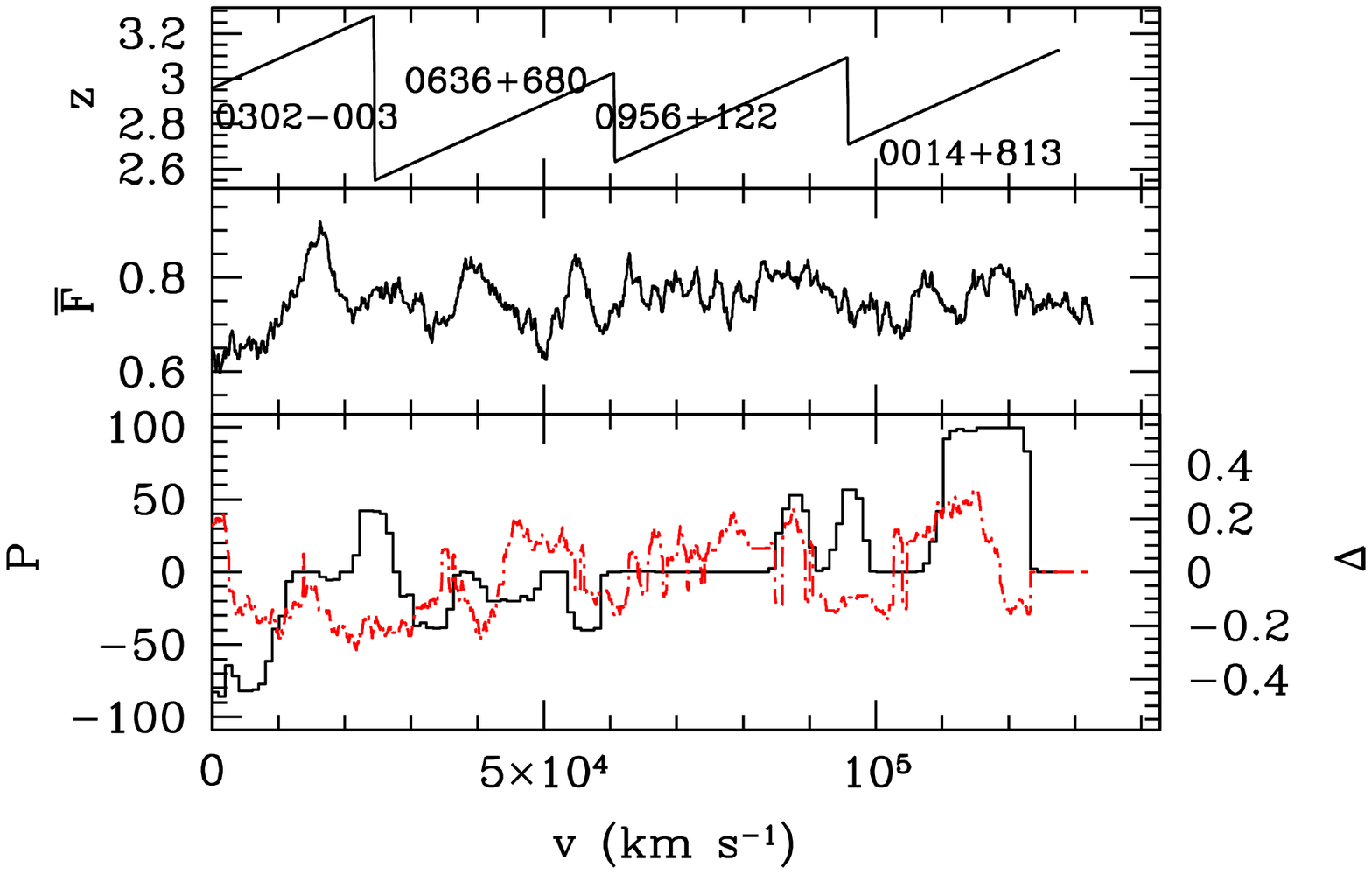}}
\end{picture}
\caption{Combined spectrum of four $\bar z\sim 2.8$ QSOs as function of
velocity $v$. Top panel: redshift range for each QSO; middle panel:
mean flux averaged over $\partial v=5\times 10^3$ km s$^{-1}$; bottom
panel: temperature indicator $\Delta(v,\partial v)$ (dot-dashed line,
right hand scale) and significance $P$ in per cent (histogram, left
hand scale). There is a significantly hotter region ($P\sim 99$ per
cent) of size $\sim 10^4$ km s$^{-1}$ at $v\sim 1.1\times 10^5$ km
s$^{-1}$ ($z\sim 3.0$) in the spectrum of QSO 0014+813.}
\label{fig:IntZ}
\end{figure}

\begin{figure}
\setlength{\unitlength}{1cm} \centering
\begin{picture}(7,7)
\put(-2., -3){\includegraphics{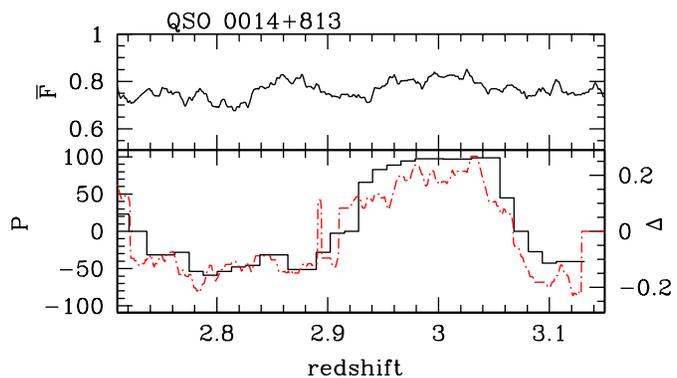}}
\end{picture}
\caption{Same as Fig.~\ref{fig:0055} but for the spectrum of QSO 0014+813,
using a window size $\partial v=5000$ km s$^{-1}$. The hot region
around $z\sim 3$ has a significance of 98.5 per cent.}
\label{fig:0014}
\end{figure}

\begin{figure*}
\setlength{\unitlength}{1cm} \centering
\begin{picture}(10,14)
\put(-3., -3.){\includegraphics{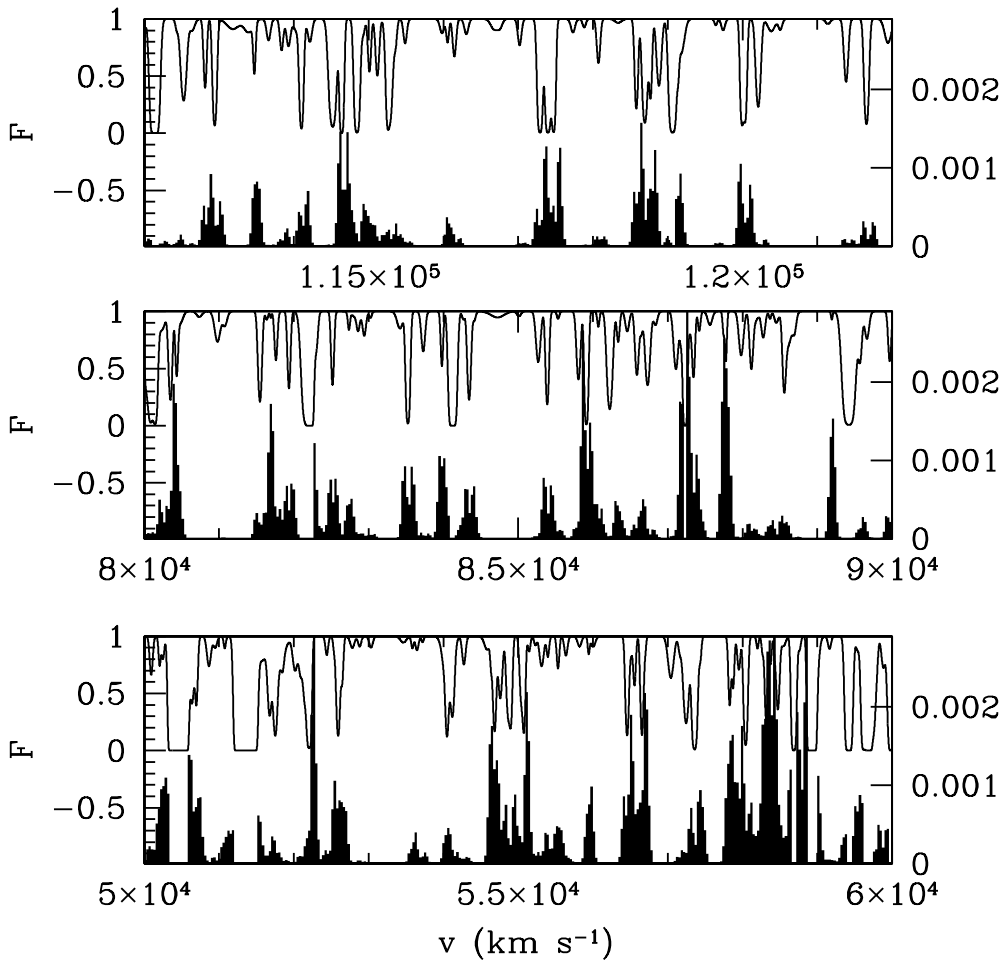}}
\end{picture}
\caption{Flux (full lines, left hand scale) and wavelet amplitudes
(filled histograms, right hand scales) for three regions of length
$10^4$ km s$^{-1}$ taken from the spectrum shown in
Fig.~\ref{fig:IntZ}. The top panel corresponds to the hot region
around $z\sim 3.1$ ($v\sim 1.1\times 10^5$ km s$^{-1}$), the middle and
bottom panel correspond to stretches of spectrum at comparable
redshifts, that are not deemed unusual. Note the small fraction of
narrow lines in the hot region, as compared to the other two.}
\label{fig:specIntZ}
\end{figure*}

We have combined the four intermediate redshift QSOs 0636+680,
0956+122, 0302--003 and 0014+813 into one longer spectrum analysed in
Fig.~\ref{fig:IntZ}.  The lowest redshift part of 0302--003 has a cold
region of moderate significance ($P\sim 85$ per cent). In addition
there is a hot region of size $\sim 10^4$ km s$^{-1}$ at a redshift
$z\sim 3.1$ in the spectrum of 0014+813. This hot region is also
significant if we analyse the spectrum of 0014+813 on its own
(Fig.~\ref{fig:0014}).

The wavelet amplitudes and the spectrum itself are compared for this
peculiar region in the spectrum of QSO 0014+813 and two other stretches
at comparable redshifts, in Fig.~\ref{fig:specIntZ}. What makes the
spectrum in the top panel peculiar from the other two, is that lines of
comparable strength are broader and consequently produce smaller
wavelet amplitudes. The analysis with 200 randomised spectra puts the
statistical significance of the hot region at the $P=99.5$ per cent
level. Using the mean wavelet amplitude to estimate the temperature, we
find that the difference in $T_0$ is 60 $\pm$ 30 per cent. This higher
temperature is presumably a consequence of the region undergoing
reionization later than the rest of the spectrum, although other
processes might contribute as well.

\subsection{Redshift range $z=[1.5,2.4]$}
\label{sect:lowz}
\begin{figure}
\setlength{\unitlength}{1cm} \centering
\begin{picture}(7,7)
\put(-1.5, -3){\includegraphics{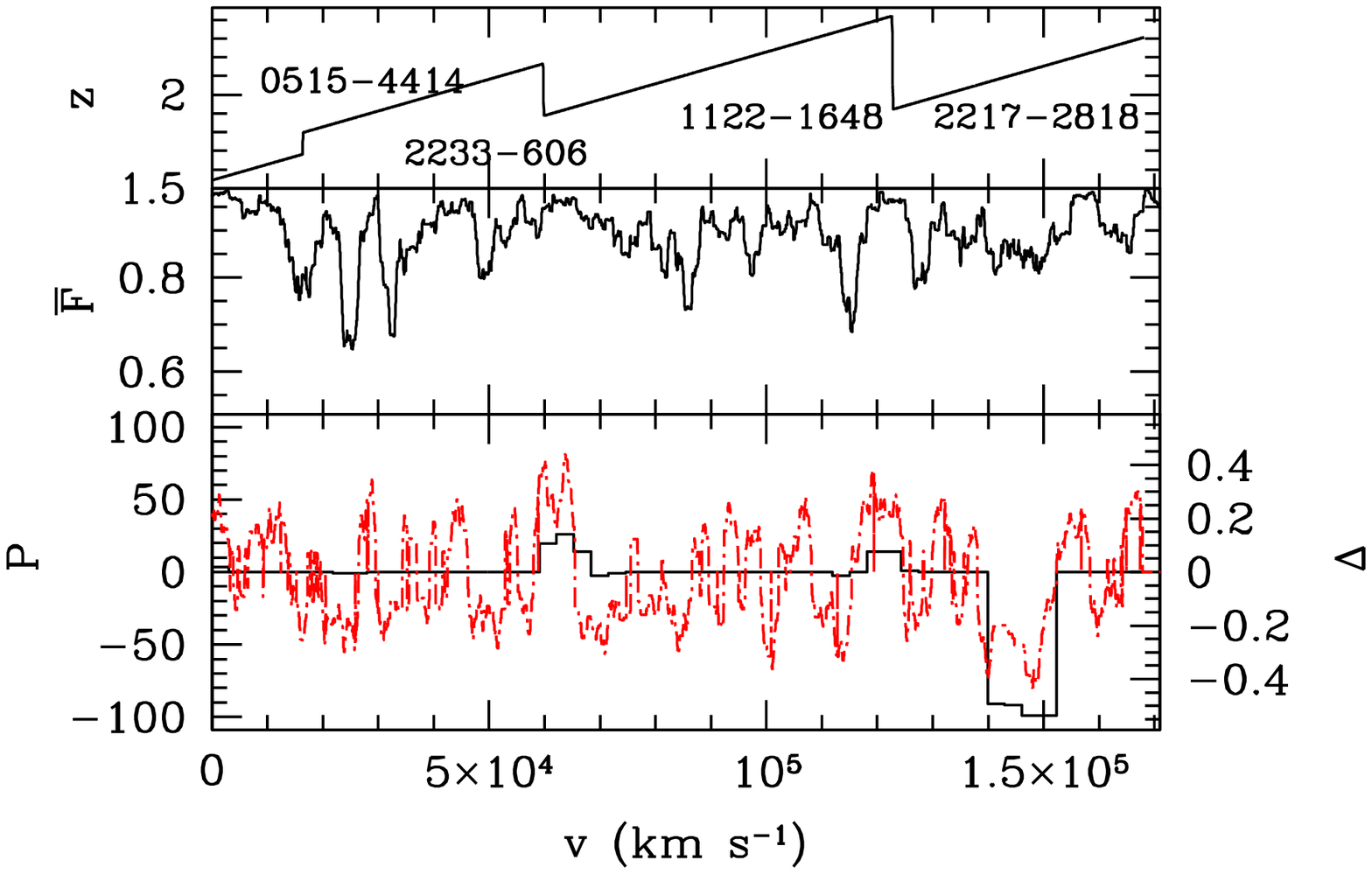}}
\end{picture}
\caption{Combined spectrum of four $\bar z\sim 2$ QSOs as function of
velocity $v$. Top panel: redshift range for each QSO; middle panel:
mean flux averaged over $\partial v=5\times 10^3$ km s$^{-1}$; bottom
panel: temperature indicator $\Delta(v,\partial v)$ (dot-dashed line,
right hand scale) and significance $P$ in per cent (histogram, left
hand scale). There is a significantly colder region ($P\sim 99$ per
cent) of size $\sim 10^4$ km s$^{-1}$ at $v\sim 1.5\times 10^5$ km
s$^{-1}$ ($z\sim 2$) in the spectrum of QSO 2217-2818.}
\label{fig:LowZ}
\end{figure}

\begin{figure}
\setlength{\unitlength}{1cm} \centering
\begin{picture}(7,7)
\put(-2., -3){\includegraphics{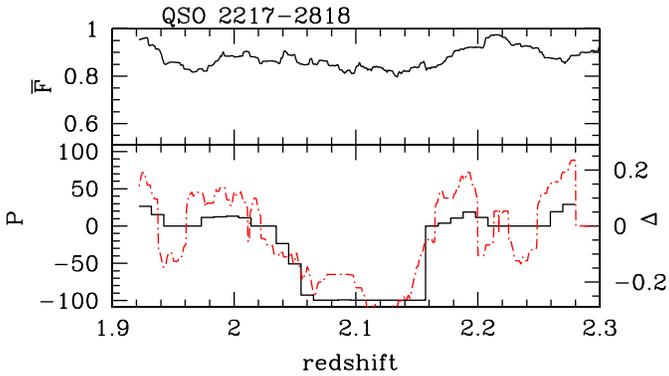}}
\end{picture}
\caption{Same as Fig.~\ref{fig:0055} but for the spectrum of the low
redshift QSO 2217-2818, using a window size $\partial v=5000$ km
s$^{-1}$. The cold region in the middle of the spectrum has a
significance of 99.5 per cent.}
\label{fig:2217}
\end{figure}

\begin{figure*}
\setlength{\unitlength}{1cm} \centering
\begin{picture}(10,14)
\put(-3., -3.){\includegraphics{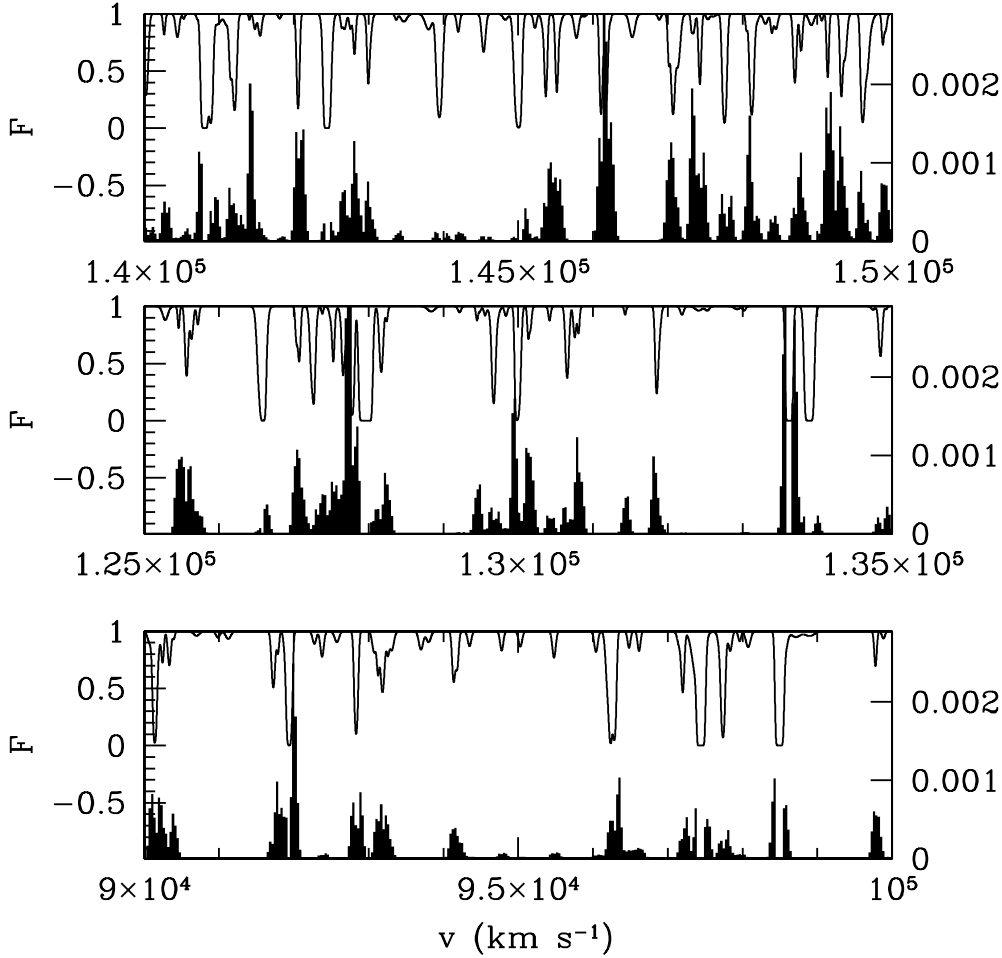}}
\end{picture}
\caption{Flux (full lines, left hand scale) and wavelet amplitudes
(filled histograms, right hand scales) for three regions of length
$10^4$ km s$^{-1}$ taken from the spectrum shown in
Fig.~\ref{fig:LowZ}. The top panel corresponds to the cold region
around $z\sim 2.1$ ($v\sim 1.5\times 10^5$ km s$^{-1}$), the middle and
bottom panel correspond to stretches of spectrum at comparable
redshifts, that are not deemed unusual. Note the large fraction of
narrow lines and the associated high wavelet amplitudes in the top
panel. The two other panels do not contain such narrow lines.}
\label{fig:specLowZ}
\end{figure*}

We have plotted the combined spectrum of the four low redshift QSOs
0515--4414 ($z=[1.5,1.7]$)], 2233-606 ($z=[1.8,2.2]$), 1122-1648
($z=[1.9,2.4]$) and 2217-2818 ($z=[1.9,2.3]$) in
Fig.~\ref{fig:LowZ}. QSO 2217-2818 contains a region of size $\sim
10^4$ km s$^{-1}$ ($\sim 100$ Mpc $h^{-1}$ comoving size in a
($\Omega_m,\Omega_\Lambda)=(0.3,0.7)$ cosmology) which is colder at the
99 per cent significance level, when compared to 200 random
realizations. The mean absorption does not appear to be unusual in this
region. When analysing the spectrum of QSO 2217-2818 on its own, the cold
region is equally significant at the 99.5 per cent level
(Fig.~\ref{fig:2217}).

The wavelet amplitudes and the spectrum itself are compared for this
cold region and two other stretches at comparable redshifts, in
Fig.~\ref{fig:specLowZ}. The differences between the top panel and the
two others, is similar to what we found for the simulated spectrum S1
in Fig.~\ref{fig:fig1}, with the narrower lines in the top panel
generating larger wavelet amplitudes. Using the mean wavelet amplitude
to measure the temperature, we find the difference in $T_0$ is 77 $\pm$
36 per cent. 

At these lower redshifts, the number of metal lines, and the fraction
of lines blended with metal lines, is no longer negligible. In order to
minimise the possibility that the detected cluster of narrow lines is
significantly contaminated by such lines, two of us have independently
fit the spectrum of QSO 2217-2818 using \vpfit. For each line not
directly identified as a metal line, we have checked whether there is a
consistent detection of Lyman-$\beta$ as well (the data are noisier at
Lyman-$\gamma$, but we also demand consistency at that transition where
possible, i.e. for $z>2.137$). Unfortunately, metal transitions often
have multiple components, and so if a transition were to be missed, it
can contribute significantly to the signal. We have performed the
wavelet analysis with these two independent line lists, and the results
are nearly identical. Both detect an unusually large wavelet signal in
the same region, at significance levels of 99.5 and 98.5 per cent
respectively.

From a theoretical point of view, if the temperature difference were
due solely to a difference in reionization redshift, then a region
could have an unusually low temperature if it had been reionised very
early, or not at all. Assuming that $T\propto (1+z)^{1.8}$ after
reionization, the early redshift would have to be $\sim 4.5$. Both
these possibilities seem unlikely to us. Given that we only detect one
such an unusual region over this redshift range, it seems more
plausible that a combination of several effects, such as residual and
partial contamination by metal lines, the presence of large scale
structure, and the gradual temperature decrease due to the expansion of
the universe, all contribute to making this region appear unusual. In
addition, the voids in the spectrum at redshifts $z\sim 1.9$, 2.2 and
2.3 (Kim et al 2001a) will contribute to the signal as well.

\section{Discussion}
Schaye et al. (2000) determined the evolution of the $\rho-T$ relation
of the IGM by measuring the cut-off in the $b-N_\H$ distribution of 9
high resolution QSOs. (Only two of those are in common with the ones
used in the present analysis). They used high resolution hydrodynamical
simulations to calibrate the relation between the $b-N_\H$ cut-off and
the underlying $\rho-T$ relation (Schaye et al. 1999). They inferred
relatively high temperatures, $T_0\sim 10^{4.1}$K, $\gamma\sim 1.3$ at
high redshifts $z=[4.5,3.5]$, an apparently sharp rise to $T_0\sim
10^{4.4}$K around $z\sim 3$ with an associated dip in $\gamma\sim 1$,
followed by a gradual decrease in $T_0$ and increase in $\gamma$
towards $z\sim 2$. This temperature evolution is not consistent with
photo-heating in the optically thin limit, as determined from the
Haardt \& Madau (1996) tracks.

Ricotti et al. (2000) wrote down a partly theoretically motivated
parametric function that describes the 2D distribution of lines in the
$b-N_\H$ plane. They employed dark matter simulations with an added \lq
pressure term\rq\ to mimic the effects of thermal smoothing, to
calibrate the fitting parameters of the function, in terms of the
simulation parameters. Their mock spectra were analysed with \autovp
(Dav\'e et al. 1997). Applying their method to published line lists,
they obtained a thermal evolution consistent with that deduced by
Schaye et al. (2000), although their error bars are large. In
particular, they also stressed the small values of $\gamma\sim 1$
around $z\sim 3$.

Bryan \& Machacek (2000) used high resolution Eulerian simulations to
measure $T_0$ from the $b-N_\H$ cut-off. These authors stressed that
the position of the cut-off also depends sensitively on the amplitude
$\sigma_8$ of the underlying dark matter power spectrum. Bryan \&
Machacek used the Voigt profile fitter described in Zhang et
al. (1997), which fits Gaussians to the {\em optical depth}
distribution of absorption lines without attempting to deblend the
absorption lines in terms of Voigt profiles, in contrast to \vpfit or
{\hbox \autovp.} Theuns et al. (2000) showed that these other fitters appear
much less sensitive to $\sigma_8$, which might partly explain the
difference in $\sigma_8$ dependence. (Theuns \& Zaroubi (2000)
demonstrated that the wavelet analysis as described here is also not
very sensitive to $\sigma_8$) Bryan \& Machacek inferred high
temperatures as well, and also concluded that the heating by \Hep might
be underestimated in the traditional models based on the optically thin
limit.

McDonald et al. (2000) used a slightly different procedure to measure
$T_0$ from the widths of the lines. They only use lines that are
well-fit by a single, unblended Voigt profile. $T_0$ is determined from
the widths of those fits, correcting for other broadening contributions
as a function of column density, using simulations. They find this
correction to be large for low density gas, and hence resort to
measuring $T_\delta$ at an overdensity of $\delta\sim 1.5$. They find
no evidence for evolution in $T_\delta\sim 10^{4.3}$K with redshift,
from $z\sim 3.9$ to 2.4. The redshift evolution of the temperature at a
given overdensity, $T_\delta$, depends on $\delta$, because the
photo-heating rate depends on density (Miralda-Escud\'e \& Rees 1994;
see also Fig.~(3) in Schaye et al. 2000). It is not clear whether the
measurement of $T_{1.5}$ by McDonald et al. (2000) is in contradiction
with the determination of $T_0$ by the other groups, given the large
uncertainty in $\gamma$ quoted by all groups. The fact that their
analysis is based on a single hydrodynamical simulation makes it
difficult to judge the robustness of their analysis procedure. In
addition, these authors averaged their results over wide redshift bins,
which hampers their ability to detect sudden changes in $T_0$ or
$\gamma$.

Zaldarriaga, Hui \& Tegmark (2000) used the small scale cut-off in the
power spectrum as a measure of $T_0$. They find similar temperatures as
McDonald et al. (2000), but unfortunately only use dark matter
simulations to calibrate their results. Also Rollinde et al. (2001) use
an inversion procedure to measure $T_0$ which is tested only on low
resolution dark matter simulations. Such simulations are inadequate to
study the thermal properties of the IGM. Even simulations that include
hydrodynamics need to be very accurate in order to resolve all the line
broadening mechanisms properly (Theuns et al 1998; Bryan et
al. 1999). Therefore this type of analysis only shows that the method
can deduce the temperature of the gas painted on top of a dark matter
simulation using an untested prescription.

All determinations so far lead to higher values of $T_0$ than were
expected for photo-heating by a UV-background dominated by QSOs, such
as computed by Haardt \& Madau (1996). One way to increase $T_0$ for a
given UV-background is to increase the baryon fraction $\omega_b\equiv
\Omega_b h^2$, since the heating rate is proportional to the physical
density of the gas (Miralda-Escud\'e \& Rees 1994). Theuns et
al. (1999) demonstrated that higher $\omega_b$ makes lines fitted by
\vpfit considerably broader as $T_0$ increases. A higher value of
$\omega_b$ would also allow galaxies to make a significant contribution
to the UV-background, as seems to be implied by the measurements of
Steidel et al. (2001) of the large escape fraction of UV-photon from
Lyman-break galaxies (Haehnelt et al. 2001). Without such a high
$\omega_b$, it would be difficult to explain the observed high opacity
of the IGM. However, such high values of $\omega_b$ would appear to
violate the tight constraints on $\omega_b\sim 0.0193\pm 0.0014$
determined from the deuterium abundance (Burles \& Tytler 1998; Kirkman
et al. 2000).

Alternatively, higher values of $T_0$ can result from other heating
mechanisms. The currently most popular one is that the calculations
based on assuming an optically thin gas underestimate the photo-heating
rate, by a factor of up to a few (Miralda-Escud\'e \& Rees 1994; Abel
\& Haehnelt 2000). Therefore, delayed \Hep reionization could be
responsible for the increase in $T_0$ around $z\sim 3$, as suggested by
Schaye et al. (2000), Ricotti et al. (2000) and Bryan \& Machacek
(2000). In fact, even the constant $T_0$ favoured by McDonald et
al. (2001) requires an extra heating source, such as \Hep reionization.

If the sources responsible for reionizing \Hep are bright but scarce,
then the temperature distribution could become quite inhomogeneous,
with large \Hepp regions separated from cooler, \Hep zones
(Miralda-Escud\'e \& Rees 1994). The ionising radiation would quickly
over run the low density voids, impulsively heating the gas, while the
more over dense filamentary regions would remain neutral for longer
(Miralda-Escud\'e, Haehnelt \& Rees 2000). Since the lower density gas
then will be hotter, its neutral fraction would become even lower
because of the $T^{-0.7}$ temperature dependence of the recombination
coefficient. As a result, the contrast in absorption between voids and
filaments would increase even further. 

The large jump in $T_0$ of 60 $\pm$ 14 per cent over the redshift
interval $z=[3.5,3.1]$ discussed in Section~\ref{sect:highz} strongly
suggests a sudden entropy injection, likely resulting from \Hep
reionization. Most of the wavelet signal comes from gas at modest over
densities $1\ltsima \delta\ltsima 3$, which produces lines of column
density $\sim 10^{13}$ cm$^{-1}$. This is because weaker lines do not
contribute significantly to the wavelets, and stronger lines are both
fewer, and partly explicitly ignored in the analysis. This gas traces
the modestly over dense filamentary pattern seen in simulations, and
our analysis suggests it becomes reionized in \Hep in a fraction $\sim
0.2$ of a Hubble time. This also naturally explains the relatively high
values of $T_0$ found at lower redshifts.

A sudden increase in $T_0$ influences the gas distribution in shallow
potential wells. Theuns et al (2000) explicitly showed how increased
heating can drive gas out of the centers of filaments into the lower
density surroundings. The velocity of this gas also contributes to the
line widths as determined by Voigt-profile fitting. Such impulsive
heating may influence the formation of halos of virial temperatures
$\ltsima 2.\times 10^4$ K (Gnedin 2000b).

After reionization, the IGM temperature will fall at a rate $T_0\approx
(1+z)^{1.8}$, slower than pure adiabatic expansion because of the
continuing photo-heating. Over a redshift extent $\delta z\sim 0.5$ of
the \lya forest of a QSO, this corresponds to a drop of $\sim$ 27 and
32 per cent at redshifts 3 and 2.5, respectively, between the high and
low redshift parts of the QSOs spectrum. This will contribute to making
the top half of QSO 0014+813 unusually hot, and the bottom half of QSO
2217-2818 unusually cold. Similarly, if \Hep reionization were delayed
in one region with respect to another, for example as a result of QSO
clustering, then we might expect to see differences in $T_0$ of order
40 per cent, given our estimate of $z=3.3\pm 0.15$ for the duration of
the reionization epoch. Such entropy fluctuations are likely to remain
present for a long time afterwards, given that the IGM evolves nearly
isentropically after reionization.

Why don't we see fluctuations on smaller scales? The analysis of
spectrum S2 suggests that the present method can detect fluctuations of
amplitude 50 per cent in $T_0$ on scales as small as $5\times 10^3$ km
s$^{-1}$. However, our simulations are clearly somewhat idealised: they
lack large scale power, and in addition have an imposed $\rho-T$
relation {\em without any} scatter. In the real universe, the presence
of large scale power and scatter in the $\rho-T$ will tend to increase
the intrinsic fluctuations in the wavelet amplitudes, making it harder
to recognise a stretch of spectrum with truly different $T_0$ or
$\gamma$. This is particularly true around reionization, where small
scale fluctuations would be hard to detect in the presence of a large
jump in the spectrum as a whole. However, the absence of small scale
fluctuations away from reionization suggests that the IGM does obey a
reasonably well defined $T-\rho$ relation on sufficiently large scales
$\ge 5000$ km s$^{-1}$. Given the arguments above it also suggest that
reionization occured relatively well synchronised throughout the IGM.

\section{Summary}
We have presented a method that uses wavelets to characterise
objectively the line widths of \lya lines in a QSO spectrum.  Colder
gas produces narrower absorption lines, for which the wavelet
amplitudes are larger. We described a statistical method, based on
recognising regions were the distribution of wavelet amplitudes is
unusual, to identify regions where the temperature at the mean density,
$T_0$, differs significantly from the mean. The significance level is
determined by performing the wavelet analysis on spectra obtained from
the data after scrambling the list of absorption lines, thereby
destroying all correlations imprinted by (temperature) fluctuations.

We applied this method to mock spectra generated from high resolution
hydrodynamical simulations. We imposed several $\rho-T$ relations on
the output from these simulations, and used these to construct spectra
with intrinsic temperature fluctuations. We demonstrated that the
method is able to detect fluctuations in $T_0$ of $\sim 50$ per cent in
regions as small as 5000 km s$^{-1}$ at redshift $z=3$, at the 95 per
cent level. We also showed that by calibrating the relation between the
average wavelet amplitude, $\langle A\rangle$, and $T_0$ using our
hydrodynamical simulations, one can estimate the IGM temperature with
an rms error of order of 30 per cent, per stretch of spectrum of 5000
km s$^{-1}$.

We have applied the method to eleven high resolution QSO spectra. There
is strong evidence of a sharp {\em increase} in $T_0$ by a factor $\sim
60\pm$ 14 per cent over the redshift interval $z=[3.5,3.1]$. We
interpreted this as evidence for \Hep reionization, and concluded that
it took of order $\delta z\sim 0.4$ to reionize the filaments at
overdensity $\sim 3$ that our method is most sensitive to. Our estimate
for the temperatures above and below the jump are $T_0\approx
10^{4.1\pm0.15}$K and $T_0\sim 10^{4.3\pm 0.15}$, respectively. These
values are in excellent agreement with those obtained by Schaye et
al. (2000), who used a very different method.

At lower redshifts, we find that the IGM appears to follow a reasonably
well defined $T-\rho$ relation on scales larger than 5000 km
s$^{-1}$. In 8 QSO spectra, we only found two unusual regions,
significant at the more than 99 per cent level. QSO 0014+813 has a hot
region of size $10^4$ km s$^{-1}$ at $z\sim 3.1$, where the temperature
is around 60$\pm$ 30 per cent higher. This could be a relic of \Hep
reionization, enhanced by the over all descrease in $T_0$ along the
spectrum as a result of adiabatic expansion. QSO 2217--2818 has a cold
region of similar size, at $z\sim 2.3$, where $T_0$ is lower by $\sim
77\pm$ 36 per cent. Again in addition to adiabatic expansion, the three
voids in this spectrum will contribute to making this region unusual.

\section*{Acknowledgements}
We would like to thank Joop Schaye for a careful reading of the paper,
and many stimulating discussions. We are also grateful to S. D'Odorico,
S. Cristiani, E. Giallongo, A. Fontana and S. Savaglio for allowing us
to use the line list of Q0055-269 prior to publication. TT thanks PPARC
for the award of an Advanced Fellowship. This work has been
supported by the \lq Formation and Evolution of Galaxies\rq\ and \lq
Physics of the Intergalactic Medium\rq\ networks set up by the European
Commission.  Research was conducted in cooperation with Silicon
Graphics/Cray Research utilising the Origin 2000 supercomputer at
DAMTP, Cambridge.

{}
\end{document}